\documentclass[preprint, superscriptaddress, showpacs,preprintnumbers,amsmath,amssymb,prb]{revtex4}
\usepackage{graphicx}

\begin{document}

\thispagestyle{empty}

\title{Thermal Casimir effect in the interaction of graphene
with dielectrics and metals}

\author{M.~Bordag}

\affiliation{Institute for Theoretical
Physics, Leipzig University, Postfach 100920,
D-04009, Leipzig, Germany}

\author{G.~L.~Klimchitskaya}

\affiliation{Institute for Theoretical
Physics, Leipzig University, Postfach 100920,
D-04009, Leipzig, Germany}

\affiliation{Central Astronomical Observatory
at Pulkovo of the Russian Academy of Sciences,
St.Petersburg, 196140, Russia}

\author{V.~M.~Mostepanenko}

\affiliation{Institute for Theoretical
Physics, Leipzig University, Postfach 100920,
D-04009, Leipzig, Germany}

\affiliation{Central Astronomical Observatory
at Pulkovo of the Russian Academy of Sciences,
St.Petersburg, 196140, Russia}

\begin{abstract}
We investigate the thermal Casimir interaction of a suspended
graphene described by the Dirac model with a plate made
of dielectric or metallic materials. The reflection coefficients
on graphene expressed in terms of a temperature-dependent polarization
tensor are used. We demonstrate that for a graphene with nonzero
mass gap parameter the Casimir free energy remains nearly
constant (and the thermal correction negligibly small) over
some temperature interval. For the interaction of graphene
with metallic
plate, the free energy is nearly the same, irrespective of
whether the metal is nonmagnetic or magnetic and
whether it is described using the Drude- or plasma-model
approaches. The free energy computed using the Dirac model
was compared with that computed using the hydrodynamic model
of graphene and big differences accessible for experimental
observation have been found. For dielectric and nonmagnetic
metallic plates described by the Drude model these differences
vanish with increasing temperature (separation). However,
for nonmagnetic metals described by the plasma model and for
magnetic metals, a severe dependence on the chosen
theoretical description of
graphene remains even at high temperature. In all cases the
analytic asymptotic expressions for the free energy at high
temperature are obtained and found in a very good agreement
with the results of numerical computations.
\end{abstract}

\maketitle

\section{Introduction}

It has been known that nanostructures based on carbon possess
unique mechanical, electrical and optical properties.\cite{1}
Among them particular attention has been given to graphene, a
two-dimensional sheet of carbon atoms arranged in a hexagonal
structure with low-energy electronic excitations described by
the Dirac equation.\cite{2} At the present time suspended graphene
membranes up to $55\,\mu$m diameter are produced.\cite{3}
This makes possible to investigate the van der Waals and Casimir
interaction between graphene and different material structures,
such as another sheet of graphene, atoms, molecules, dielectric
and metallic plates, spheres etc. Respective theoretical
investigations were performed using the phenomenological
density-functional methods,\cite{4,5,6,7,8} second-order
perturbation theory\cite{9} and, for multilayered carbon
nanostructures, using the Lifshitz theory.\cite{10}

To apply the Lifshitz theory, one needs the reflection
coefficients on graphene over a wide frequency region.
It is customary to express the reflection coefficients in terms
of the frequency-dependent dielectric permittivity, a concept
which is not well defined for one-atom-thick carbon
nanostructures.
Because of this, two models for the reflection coefficients
on graphene with no use of dielectric permittivity were
proposed, the hydrodynamic one\cite{11,12} and the Dirac
one.\cite{13,14} In the framework of the hydrodynamic model,
graphene is considered as an infinitesimally thin positively
charged sheet, carrying a homogeneous fluid with some mass and
negative charge densities. The hydrodynamic model was applied
for calculation of the Casimir and Casimir-Polder
interactions.\cite{15,16,17} In the framework of the Dirac
model, it is taken into account that for energies below a few
eV the dispersion relation for quasiparticles in graphene is
linear with respect to the momentum, whereas it is quadratic
in the hydrodynamic model. The reflection
coefficients of graphene at zero and nonzero temperature
were found in Refs.~\cite{18} and{\ }\cite{19}, respectively.
Some calculations of the Casimir-Polder graphene-atom
interaction were performed at zero temperature using both
the hydrodynamic and Dirac models.\cite{20,21}
It should be remembered, however, that both these models
are only approximations. As was already mentioned, the
hydrodynamic model disregards the Dirac character of charge
carriers in graphene. As to the Dirac model, it extends the
linear dispersion relation for quasiparticles to any energy,
whereas this property applies only at low energies.
Because of this, one can conclude that in calculations of
the Casimir and Casimir-Polder forces the Dirac model of
graphene should be applicable at large separations between
the test bodies, whereas the hydrodynamic model might work
at short separation distances.

There is also another approach to the application of the
Lifshitz theory to graphene. In this approach graphene is
characterized by a spatially nonlocal dielectric permittivity
depending both on the frequency and the wave vector.
Different authors express such a dielectric permittivity
either through the polarizability of one graphene
layer \cite{21a} or in the random phase approximation.\cite{21b}

It is well known that thermal Casimir effect is a subject of
debate and there are different theoretical approaches to its
description.\cite{22,23,24} For two graphene sheets
in a nonretarded regime it was found\cite{25} that a relatively
large thermal correction to the Casimir force at room temperature
arises at short separation distances of tens of nanometers.
This conclusion was qualitatively confirmed,\cite{19} using
the Dirac model of graphene, with an alternative explanation.
The reason for the origin of large thermal correction for
graphene is that the contribution of all terms with nonzero
Matsubara frequencies at room temperature becomes small in
comparison with the zero-frequency term even at short
separations. It was also found\cite{26} that in graphene-atom
Casimir-Polder interaction the thermal effect depends crucially
on the magnitude of a mass gap parameter in the Dirac model.
Specifically, for a nonzero gap there exists an interval of
temperatures (separations) where the thermal correction remains
small with increasing temperature (separation).
The possibility of large thermal correction at short separations
links the Dirac model of graphene to the Drude model used in the literature
to describe the Casimir effect between real metals.\cite{23,24}
Because of this, it is of much interest to investigate the thermal
Casimir effect in the interaction of graphene with real material
bodies made of different materials.

In this paper, we calculate the free energy of the Casimir
interaction between a suspended  graphene membrane described by
the Dirac model and dielectric (silicon, sapphire) or metallic
(Au, Ni) plates. In so doing materials of the plate are described
by realistic dielectric permittivities taking into account the
interband transitions of core electrons. We demonstrate that,
similar to graphene-atom interaction, the behavior of thermal
correction crucially depends on the mass gap parameter $\Delta$
of the Dirac model.
Note that although the Dirac-type excitations in pristine
graphene are gapless, the influence of electron-electron
interaction,
substrates, defects of structure, and other effects leads to
a nonzero mass gap.\cite{2,26a,26b,26c,26d}
Specifically, we show that larger is the
magnitude of mass gap parameter, wider is the separation
(temperature) interval, where the thermal correction remains
small with increasing separation (temperature).
For a metallic plate interacting with graphene, we perform all
calculations using both the Drude- and plasma-model approaches
to the dielectric permittivity of metal.
In the case when graphene described by the Dirac model interacts
with a metallic plate (either nonmagnetic or magnetic) the
calculation results obtained using both approaches nearly
coincide and do not depend on the magnetic properties.
In contrast to the case of two Drude metals, the thermal
correction for a graphene-metal interaction has the same sign as
the interaction energy at zero temperature, i.e., the magnitude
of the Casimir free energy increases with increasing temperature.
Note that all results obtained for a free energy in graphene-plate
geometry can be reformulated as the Casimir force
between a material sphere and
a graphene sheet using the proximity force approximation
(PFA).\cite{27} As was recently shown,\cite{28,28a,29}
the error arising from the use of PFA is less than the ratio of
separation distance to sphere radius.

In this paper we also compare the predictions of the Dirac model
for
the thermal Casimir effect with respective predictions of the
hydrodynamic model and discuss the application region of each.
Specifically, it is shown that the hydrodynamic and Dirac models
of graphene lead to different results at short separations and
to nearly coinciding results at large separations for the
Casimir free energy of graphene interacting at room temperature
with a dielectric plate or with a nonmagnetic metallic plate
described by the Drude model. For a nonmagnetic metallic plate
described by the plasma model or for a magnetic plate the
predictions of the hydrodynamic and Dirac models of graphene
are significantly different at all separations considered and
can be discriminated experimentally. At large separations the
asymptotic expressions for the Casimir free energy of graphene-plate
interaction are derived and compared with the
computational results.

The paper is organized as follows. In Sec.~II we present the
reflection coefficients of the electromagnetic oscillations on
graphene in the Dirac and hydrodynamic models. Section~III is
devoted to the thermal Casimir interaction of graphene described
by the Dirac model with a dielectric plate made of silicon or
sapphire. Similar results for graphene interacting with
a metallic plate made of Au and Ni
are presented in Sec.~IV. In Sec.~V the
theoretical predictions following from the Dirac and
hydrodynamic models are compared. Our conclusions and
discussions are contained in Sec.~VI.

\section{Reflection coefficients on graphene}

Here, we briefly present the Lifshitz formula for the free
energy of graphene interacting with a material plate and
respective reflection coefficients on graphene derived using
the Dirac and hydrodynamic models. It is supposed that
the suspended graphene is at a separation $a$ from the thick plate
(semispace) at thermal equilibrium at temperature $T$.
The material of a plate is described by the frequency-dependent
dielectric permittivity $\varepsilon(\omega)$ and magnetic
permeability $\mu(\omega)$. The Casimir free energy per unit
area ${\cal F}$ is given by the Lifshitz formula.\cite{27}
For simplicity in computations, we express it in terms of
dimensionless variables as follows:
\begin{eqnarray}
&&
{\cal F}(a,T)=\frac{k_BT}{8\pi a^2}\sum_{l=0}^{\infty}
{\vphantom{\sum}}^{\prime}\int_{\zeta_l}^{\infty}y\,dy
\nonumber \\
&&~~~~\times
\left\{\ln\left[1-r_{\rm TM}^{(g)}(i\zeta_l,y)
r_{\rm TM}^{(p)}(i\zeta_l,y)e^{-y}\right]\right.
\nonumber \\
&&~~~~~~
\left.+\ln\left[1-r_{\rm TE}^{(g)}(i\zeta_l,y)
r_{\rm TE}^{(p)}(i\zeta_l,y)e^{-y}\right]\right\}.
\label{eq1}
\end{eqnarray}
\noindent
Here, $k_B$ is the Boltzmann constant,
$\zeta_l$ are the dimensionless Matsubara frequencies connected
with the dimensional ones, $\xi_l=2\pi k_BTl/\hbar$, by the
equality $\zeta_l=\xi_l/\omega_c$ where $\omega_c=c/(2a)$.
The dimensionless wave vector variable $y$ is connected with the
magnitude of the projection of the wave vector on the plane of a
plate, $k_{\bot}$, by the equality
$y=2a(k_{\bot}^2+\xi_l^2/c^2)^{1/2}$.
The reflection coefficients on graphene, $r_{\rm TM(TE)}^{(g)}$,
and on a plate, $r_{\rm TM(TE)}^{(p)}$, are for two independent
polarizations of the electromagnetic field, transverse magnetic
(TM) and transverse electric (TE). They are taken at
imaginary frequencies. The prime near the summation sign means
that the term with $l=0$ is taken with a factor 1/2.

The reflection coefficients on graphene in the Dirac model
were expressed in terms of the components of the polarization
tensor in three-dimensional space-time in the following
way\cite{18,19}
\begin{eqnarray}
&&
r_{\rm TM}^{(g)}(i\zeta_l,y)=
\frac{y\tilde{\Pi}_{00}}{y\tilde{\Pi}_{00}+
2(y^2-\zeta_l^2)},
\label{eq2} \\
&&
r_{\rm TE}^{(g)}(i\zeta_l,y)=
-\frac{(y^2-\zeta_l^2)\tilde{\Pi}_{tr}-
y^2\tilde{\Pi}_{00}}{(y^2-\zeta_l^2)(\tilde{\Pi}_{tr}
+2y)-y^2\tilde{\Pi}_{00}}.
\nonumber
\end{eqnarray}
\noindent
Here, the dimensionless components of the polarization tensor
are defined as
\begin{equation}
\tilde{\Pi}_{00,tr}=\frac{2a}{\hbar}{\Pi}_{00,tr}
\label{eq3}
\end{equation}
\noindent
and trace stands for the sum of spatial components
$\Pi_{1}^{\,1}$ and $\Pi_{2}^{\,2}$.

At nonzero temperature the explicit expression for the
polarization tensor in the Dirac model with arbitrary mass gap
parameter $\Delta$ and chemical potential $\mu$ was found in
Ref.~\cite{19}. Here we consider the case of undoped graphene
and put $\mu=0$. Then in terms of our dimensionless variables
the result of Ref.~\cite{19} for the 00-component of the
polarization tensor takes the following equivalent form:\cite{26}
\begin{eqnarray}
&&
\tilde{\Pi}_{00}(i\zeta_l,y)=8\alpha(y^2-\zeta_l^2)
\int_{0}^{1}dx\frac{x(1-x)}{\left[{\tilde{\Delta}}^2+
x(1-x)f(\zeta_l,y)\right]^{1/2}}
+\frac{8\alpha}{{\tilde{v}}_F^2}\int_{0}^{1}dx
\label{eq4} \\
&&
~\times
\left\{\vphantom{\frac{{\tilde{\Delta}}^2+\zeta_l^2x(1-x)}{\left[{\tilde{\Delta}}^2+
x(1-x)f(\zeta_l,y)\right]^{1/2}}}
\frac{\tau}{2\pi}\ln\left[1+2\cos(2\pi lx)e^{-g(\tau,\zeta_l,y)}
+e^{-2g(\tau,\zeta_l,y)}\right]
-\frac{\zeta_l}{2}(1-2x)
\frac{\sin(2\pi lx)}{\cosh{g(\tau,\zeta_l,y)}+
\cos(2\pi lx)}
\right.
\nonumber \\
&&~
\left.
+\frac{{\tilde{\Delta}}^2+\zeta_l^2x(1-x)}{\left[{\tilde{\Delta}}^2+
x(1-x)f(\zeta_l,y)\right]^{1/2}}\,
\frac{\cos(2\pi lx)+e^{-g(\tau,\zeta_l,y)}}{\cosh{g(\tau,\zeta_l,y)}+
\cos(2\pi lx)}
\right\}.
\nonumber
\end{eqnarray}
\noindent
In this expression, $\alpha=e^2/(\hbar c)$ in the fine-structure
constant ($e$ is the electron charge),
$\tilde{\Delta}=\Delta/(\hbar\omega_c)$ is the
dimensionless mass gap parameter, the dimensionless Fermi velocity
is $\tilde{v}_F=v_F/c\sim 1/300$, and the dimensionless variable $\tau$ is
defined as $\tau=2\pi T/T_{\rm eff}=4\pi ak_BT/(\hbar c)$.
Equation (\ref{eq4}) also contains two dimensionless functions
defined by
\begin{eqnarray}
&&
f(\zeta_l,y)={\tilde{v}}_F^2y^2+(1-{\tilde{v}}_F^2)\zeta_l^2,
\label{eq5} \\
&&
g(\tau,\zeta_l,y)=\frac{2\pi}{\tau}
\left[{\tilde{\Delta}}^2+x(1-x)f(\zeta_l,y)\right]^{1/2}.
\nonumber
\end{eqnarray}
\noindent
The result of Ref.~\cite{19} for the trace of the polarization
tensor in terms of our dimensionless variables is given
by\cite{26}
\begin{eqnarray}
&&
\tilde{\Pi}_{tr}(i\zeta_l,y)=8\alpha[y^2+f(\zeta_l,y)]
\int_{0}^{1}dx\frac{x(1-x)}{\left[{\tilde{\Delta}}^2+
x(1-x)f(\zeta_l,y)\right]^{1/2}}
+\frac{8\alpha}{{\tilde{v}}_F^2}\int_{0}^{1}dx
\label{eq6} \\
&&
~\times
\left\{\vphantom{\frac{{\tilde{\Delta}}^2+\zeta_l^2x(1-x)}{\left[{\tilde{\Delta}}^2+
x(1-x)f(\zeta_l,y)\right]^{1/2}}}
\frac{\tau}{2\pi}\ln\left[1+2\cos(2\pi lx)e^{-g(\tau,\zeta_l,y)}
+e^{-2g(\tau,\zeta_l,y)}\right]
\right.
\nonumber \\
&&~
-\frac{\zeta_l(1-2{\tilde{v}}_F^2)}{2}(1-2x)
\frac{\sin(2\pi lx)}{\cosh{g(\tau,\zeta_l,y)}+
\cos(2\pi lx)}
\nonumber \\
&&~
\left.
+\frac{{\tilde{\Delta}}^2+x(1-x)[(1-{\tilde{v}}_F^2)^2\zeta_l^2-
{\tilde{v}}_F^4y^2]}{\left[{\tilde{\Delta}}^2+
x(1-x)f(\zeta_l,y)\right]^{1/2}}\,
\frac{\cos(2\pi lx)+e^{-g(\tau,\zeta_l,y)}}{\cosh{g(\tau,\zeta_l,y)}+
\cos(2\pi lx)}
\right\}.
\nonumber
\end{eqnarray}
\noindent
The properties of the reflection coefficients (\ref{eq2}) with the
polarization tensor (\ref{eq4}) and (\ref{eq6}) were studied
previously.\cite{19,26}

In the framework of the hydrodynamic model discussed in Sec.~I the
reflection coefficients on graphene take the more simple
form\cite{11,12,15,16}
\begin{eqnarray}
&&
r_{\rm TM}^{(g)}(i\zeta_l,y)=
\frac{\tilde{K}y}{\tilde{K}y+\zeta_l^2},
\nonumber \\
&&
r_{\rm TE}^{(g)}(i\zeta_l,y)=
-\frac{\tilde{K}}{\tilde{K}+y}.
\label{eq7}
\end{eqnarray}
\noindent
Here, the dimensionless characteristic wave number of the graphene
is defined as $\tilde{K}=2aK$, where the dimensional wave number is
\begin{equation}
K=2\pi\frac{ne^2}{mc^2}=6.75\times 10^{5}\,\mbox{m}^{-1}.
\label{eq8}
\end{equation}
\noindent
In Eq.~(\ref{eq8}) $n$ is the number of $\pi$-electrons per unit
area, $m$ is the electron mass. Note that the parameter $K$ of the
hydrodynamic model does not depend on temperature. Thus, the
reflection coefficients (\ref{eq7})
and the free energy (\ref{eq1}) in the case of hydrodynamic model
depend on the temperature
only through the Matsubara frequencies. A different situation
arises in the Dirac model. Here, the reflection coefficients
(\ref{eq2}) depend on $T$ not only through the Matsubara
frequencies, but also through the components of the polarization
tensor. This links the Dirac model of graphene to the
Drude-model approach to the thermal Casimir force between real
metals (because the relaxation parameter of the Drude model and,
thus, the reflection coefficients are the explicit functions of
temperature). In this respect the use of the hydrodynamic model
of graphene in calculations of the Casimir force is analogous
to the plasma-model approach where the reflection coefficients
depend on $T$ only through the Matsubara frequencies.

As to the reflection coefficient of electromagnetic oscillations
on thick metallic plate (semispace), they have the standard
form\cite{27}
\begin{eqnarray}
&&
r_{\rm TM}^{(p)}(i\zeta_l,y)=
\frac{\varepsilon_l y-\left[y^2+\zeta_l^2(\varepsilon_l\mu_l-
1)\right]^{1/2}}{\varepsilon_l y+\left[y^2+\zeta_l^2(\varepsilon_l\mu_l-
1)\right]^{1/2}},
\nonumber \\
&&
r_{\rm TE}^{(p)}(i\zeta_l,y)=
\frac{\mu_l y-\left[y^2+\zeta_l^2(\varepsilon_l\mu_l-
1)\right]^{1/2}}{\mu_l y+\left[y^2+\zeta_l^2(\varepsilon_l\mu_l-
1)\right]^{1/2}}.
\label{eq9}
\end{eqnarray}
\noindent
Here, both the dielectric permittivity
$\varepsilon_l\equiv\varepsilon(i\omega_c\zeta_l)$ and the
magnetic permeability $\mu_l\equiv\mu(i\omega_c\zeta_l)$
are calculated along the imaginary frequency axis.

\section{Thermal interaction of graphene described by the Dirac
model with dielectric plate}
Here, we investigate dependences of the free energy as functions
of temperature and separation in the Casimir interaction of
a suspended graphene sheet with the dielectric plate made of
different dielectrics. As the most typical dielectric
materials we consider silicon (Si) and sapphire (Al${}_2$O${}_3$)
which possess relatively large values of the static dielectric
permittivity [$\varepsilon(0)=11.67$
and 10.102, respectively] but quite different
behaviors along the axis of imaginary frequencies (for Si it is
caused by electronic polarization, whereas for Al${}_2$O${}_3$
by both electronic and ionic polarizations).

\subsection{Free energy as a function of temperature}

We begin with the free energy of interaction of graphene described
by the Dirac model with Si plate. The dielectric permittivity of
Si along the imaginary frequency axis was obtained\cite{30,31} by
means of the Kramers-Kronig relations from the tabulated optical data
for the complex index of refraction.\cite{32}
It is assumed to be temperature-independent.
Computations were performed by Eq.~(\ref{eq1}), where the
reflection coefficients on graphene are given by Eqs.~(\ref{eq2}),
(\ref{eq4}) and (\ref{eq6}), and on silicon by Eq.~(\ref{eq9})
with $\mu_l=1$ and $\varepsilon_l$ specified above, over the
temperature interval from 0 to 300\,K at two separation distances
$a=100\,$nm and $a=1\,\mu$m.
For the mass gap parameter of the Dirac model of graphene
only the upper bound is known.\cite{18,26b,26d}
For a suspended graphene we choose the realistic upper bound
$\Delta\leq 0.1\,$eV.
For a graphene deposited on substrate, $\Delta$ can be several
times larger.\cite{26b} Taking this upper bound into account,
 we perform all
computations for $\Delta=0.1$, 0.05, 0.01, and 0\,eV.

\begin{figure*}[h]
\vspace*{-1.0cm}
\centerline{\hspace*{-1cm}
\includegraphics{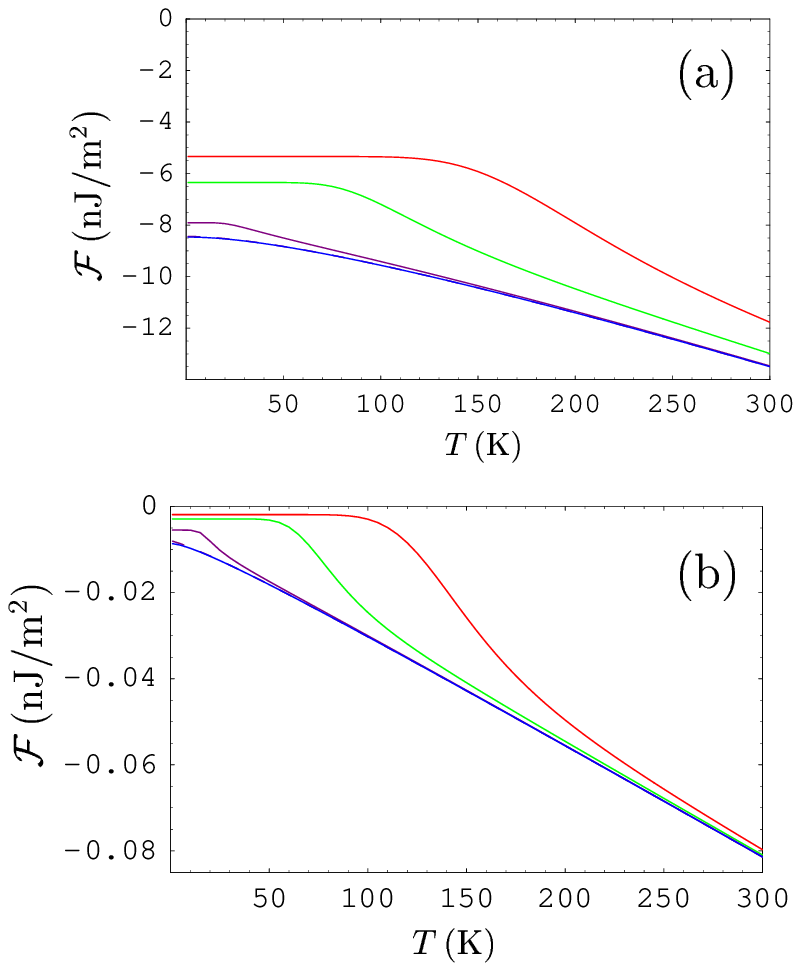}
}
\vspace*{-16.cm}
\caption{\label{fg1}(Color online)
The Casimir free energy per unit area as a function of temperature
for the interaction of
graphene described by the Dirac model with Si plate
(a) at $a=100\,$nm and (b) at $a=1\,\mu$m.
The lines from top to bottom correspond to the mass gap
parameter $\Delta=0.1$, 0.05, 0.01, and $\lesssim 0.001\,$eV,
respectively.
}
\end{figure*}

The computational results for the free energy per unit area are
presented in Fig.~\ref{fg1} as functions of temperature, where
the lines from top to bottom correspond to $\Delta$ decreasing
from 0.1\,eV to 0\,eV, (a) at the separation $a=100\,$nm and
(b) at $a=1\,\mu$m.
As can be seen in Fig.~\ref{fg1}(a,b), for each mass gap
$\Delta\neq 0$ there exists the temperature interval where the
free energy remains nearly constant with increasing temperature.
The width of these intervals quickly decreases with decreasing
$\Delta$. Thus, at $a=100\,$nm and $\Delta=0.1\,$eV the free
energy is nearly constant up to $T=120\,$K. At the same
separation,
but with $\Delta=0.05\,$eV, the same property holds only up
to $T=70\,$K. The widths of intervals, where the free energy
remains nearly constant with increasing $T$, are also narrowed
with the increase of separation [see Fig.~\ref{fg1}(b)].
Note that computations performed for $\Delta\lesssim 0.001\,$eV
lead to nearly the same numerical results as for $\Delta=0$
[the lowest lines in Fig.~\ref{fg1}(a,b)].

Figure~\ref{fg1} suggests that within  the temperature interval,
where the free energy is nearly flat, the thermal correction to
the Casimir energy at zero temperature should be relatively small.
We confirm this conclusion by the direct computation of the thermal
correction to the Casimir energy defined as
\begin{equation}
\Delta_T{\cal F}(a,T)={\cal F}(a,T)-{\cal F}(a,0)
\label{eq10}
\end{equation}
\noindent
for the same values of parameters, as in Fig.~\ref{fg1}.

\begin{figure*}[h]
\vspace*{-1.0cm}
\centerline{\hspace*{2.0cm}
\includegraphics{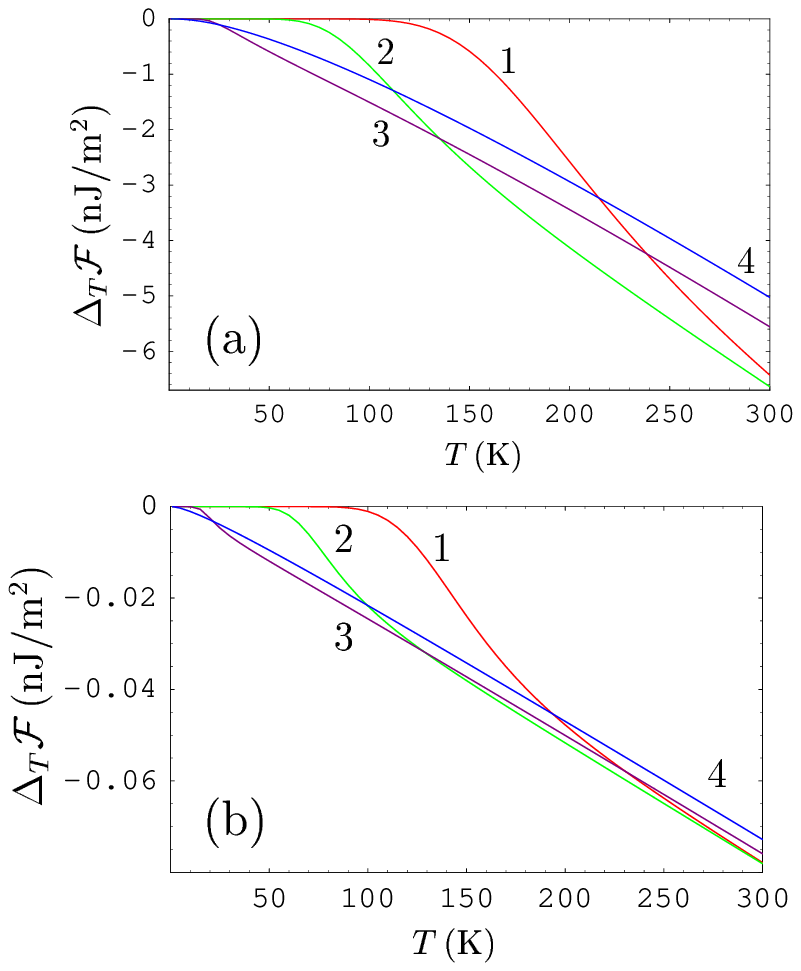}
}
\vspace*{-16.cm}
\caption{\label{fg2}(Color online)
The thermal correction to the Casimir energy per unit area
at zero temperature as a function of temperature
for the interaction of
graphene described by the Dirac model with Si plate
(a) at $a=100\,$nm and (b) at $a=1\,\mu$m.
The lines labeled 1, 2, 3, and 4 correspond to the mass gap
parameter $\Delta=0.1$, 0.05, 0.01, and $\lesssim 0.001\,$eV,
respectively.}
\end{figure*}

The computational results for the thermal correction as a function
of temperature are presented in Fig.~\ref{fg2} at separations
(a) $a=100\,$nm and (b) $a=1\,\mu$m. The lines labeled 1, 2, 3,
and 4 correspond to the values of mass gap parameter $\Delta=0.1$,
0.05, 0.01, and $\lesssim 0.001\,$eV, respectively.
As expected, for each line with $\Delta\neq 0$ there is some
interval where the thermal correction remains nearly zero.
These intervals are just the same where the Casimir free energy
in Fig.~\ref{fg1} remains nearly flat. It is interesting
to note that the
thermal corrections are monotonously decreasing functions of
temperature and have the same negative sign as the free energy.
This contrasts with the Drude model approach to the Casimir force
between real metals where the thermal correction over a wide
temperature interval is positive making the free energy the
nonmonotonous function of temperature.\cite{23,27}

\begin{figure*}[h]
\vspace*{-1.0cm}
\centerline{\hspace*{-1cm}
\includegraphics{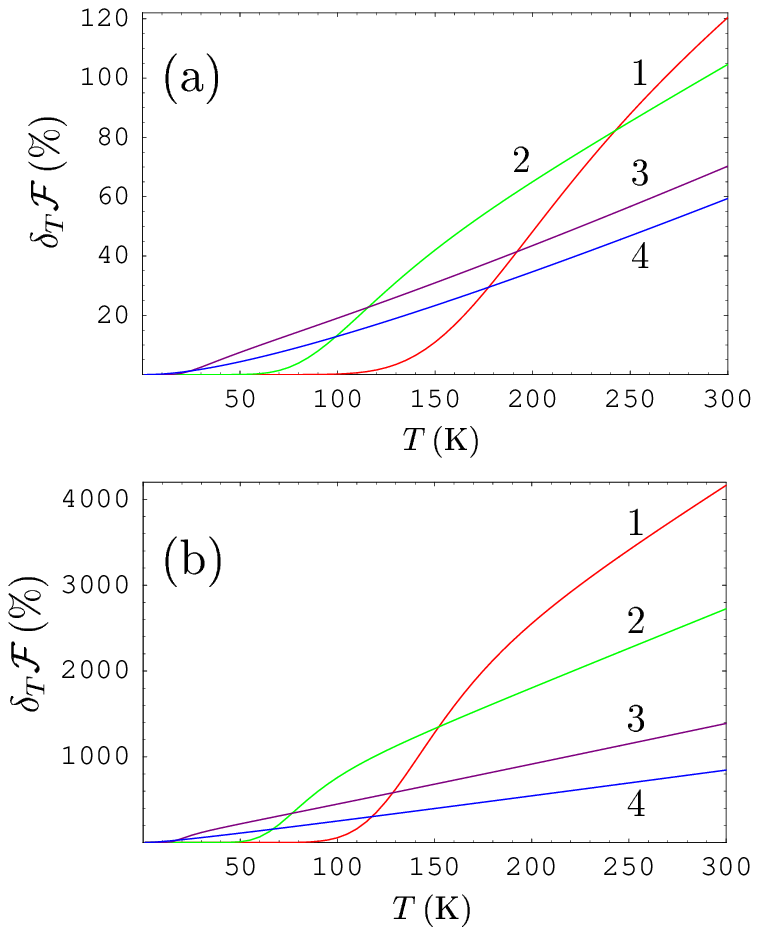}
}
\vspace*{-16.cm}
\caption{\label{fg3}(Color online)
The relative thermal correction to the Casimir energy
at zero temperature as a function of temperature
for the interaction of
graphene described by the Dirac model with Si plate
(a) at $a=100\,$nm and (b) at $a=1\,\mu$m.
The lines labeled 1, 2, 3, and 4 correspond to the mass gap
parameter $\Delta=0.1$, 0.05, 0.01, and $\lesssim 0.001\,$eV,
respectively.
}
\end{figure*}

It is instructive also to compute the relative thermal correction
to the Casimir energy defined as
\begin{equation}
\delta_T{\cal F}(a,T)=\frac{\Delta_T{\cal F}(a,T)}{{\cal F}(a,0)}.
\label{eq11}
\end{equation}
\noindent
The computational results are presented in Fig.~\ref{fg3} in
percent as functions of temperature at separations distances
(a) $a=100\,$nm and (b) $a=1\,\mu$m. The lines 1, 2, 3,
and 4 are labeled in the same way as in Fig.~\ref{fg2}.
As can be seen in Fig.~\ref{fg3}, all the relative thermal
corrections are monotonously increasing functions of temperature
whose character depends crucially on the magnitude of a mass
gap parameter $\Delta$. For example, at $a=100\,$nm
[Fig.~\ref{fg3}(a)] the relative thermal correction of line 1
($\Delta=0.1\,$eV) is nearly equal to zero below $T=120\,$K
but achieves 120.44\% at $T=300\,$K. To compare, at the same
separation the relative thermal correction of line 4
($\Delta\lesssim 0.001\,$eV), which is much larger than the
corrections of lines 1--3 at low temperatures, achieves only
59.40\% at $T=300\,$K. It is seen that at room temperature
the relative thermal correction in the Casimir interaction
of graphene with Si is rather large even at relatively short
separations. From Fig.~\ref{fg3}(b) it follows that at
$a=1\,\mu$m the relative thermal correction at room
temperature exceeds 4000\%, i.e., the absolute thermal
correction exceeds the Casimir energy at zero temperature
by a factor of 40. This makes graphene interesting for
experimental investigation of thermal effects in the
Casimir force.

\begin{figure*}[h]
\vspace*{-1.0cm}
\centerline{\hspace*{2.0cm}
\includegraphics{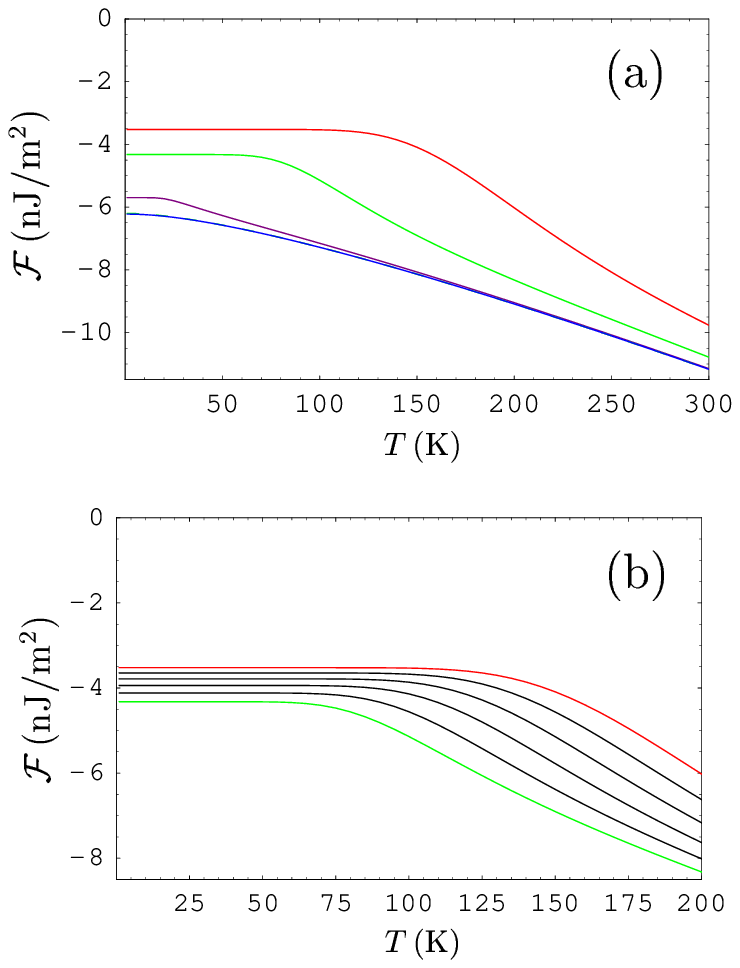}
}
\vspace*{-16.cm}
\caption{\label{fg4}(Color online)
The Casimir free energy per unit area as a function of temperature
at $a=100\,$nm for the interaction of
graphene described by the Dirac model with sapphire plate.
The lines from top to bottom correspond to the mass gap
parameter (a) $\Delta=0.1$, 0.05, 0.01, and $\lesssim 0.001\,$eV
and (b) $\Delta=0.1$, 0.09, 0.08, 0.07, 0.06, and 0.05\,eV,
respectively.
}
\end{figure*}

Next we consider the interaction of graphene described by the
Dirac model with sapphire plate. The dielectric permittivity
of sapphire along the imaginary frequency axis is well
described\cite{33} in the Ninham-Parsegian approximation
and was already used\cite{31} in computations of the Casimir
force. Computations of the free energy as a function of
temperature were performed in the same way as for silicon.
The computational results for $a=100\,$nm are presented in
Fig.~\ref{fg4}(a) by the four lines from top to bottom for
$\Delta=0.1$, 0.05, 0.01, and $\lesssim 0.001\,$eV,
respectively.
Similar to Fig.~\ref{fg1}(a), there are temperature intervals
where the free energy remains nearly constant. For sapphire,
however, the respective magnitudes of the free energy are
smaller than for a silicon. Skipping the computational
results at $a=1\,\mu$m [which are similar to those presented
in Fig.~\ref{fg1}(b)], we present in Fig.~\ref{fg4}(b) the
more detailed computational results for the free energy in the
temperature interval from 0\,K to 200\,K, where the lines
from top to bottom correspond to $\Delta=0.1$, 0.09, 0.08, 0.07,
0.06, and 0.05\,eV, respectively. Keeping in mind that the
exact value of $\Delta$ is not known, computations of this
kind can be useful for the determination of $\Delta$ from
the comparison between experiment and theory.
For sapphire, the computational results for the absolute and
relative thermal corrections are similar to those for silicon
(see Figs.~\ref{fg2} and \ref{fg3}). Because of this we do not
present them here.

\subsection{Free energy as a function of separation}

Here, we present the computational results for the interaction of
graphene described by the Dirac model with dielectric plates
as a function of separation. The same equations and dielectric
functions, as  in Sec.~IIIA, are used. Taking into account that
the Casimir free energy strongly depends on separation, we
normalize the results obtained on the Casimir energy per unit area
between two parallel plates made of ideal metal
\begin{equation}
E_C(a)=-\frac{\pi^2}{720}\,\frac{\hbar c}{a^3}.
\label{eq12}
\end{equation}

\begin{figure*}[h]
\vspace*{-1.0cm}
\centerline{\hspace*{-1cm}
\includegraphics{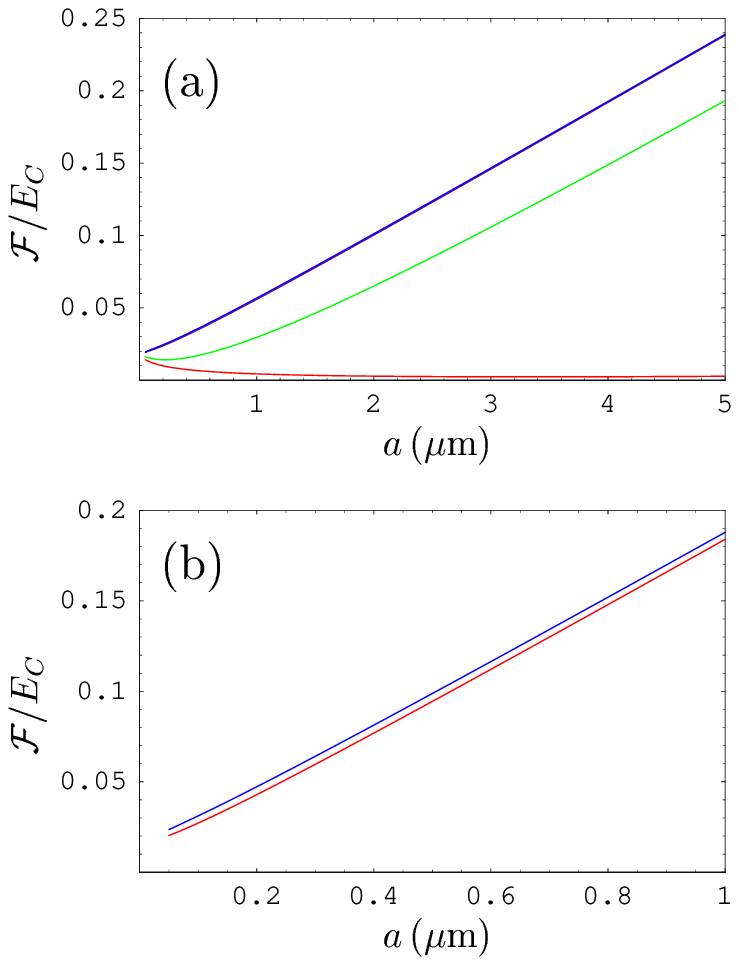}
}
\vspace*{-16.cm}
\caption{\label{fg5}(Color online)
The Casimir free energy  as a function of separation
for the interaction of
graphene described by the Dirac model with Si plate
(a) at $T=77\,$K and (b) at $T=300\,$K
normalized on the Casimir energy between two parallel ideal
metal planes.
The lines from  bottom to top correspond to the mass gap
parameter (a) $\Delta=0.1$, 0.05,  $\lesssim 0.01\,$eV
and (b) $\Delta=0.1$, $\lesssim 0.01\,$eV,
respectively.
}
\end{figure*}

In Fig.~\ref{fg5}(a) the quantity ${\cal F}/E_C$ at $T=77\,$K
is plotted as a function of separation over the region from
50\,nm to $5\,\mu$m. The lines from bottom to top correspond
to the mass gap parameter equal to 0.1, 0.05, and
$\lesssim 0.01\,$eV, respectively. In all cases the magnitude
of the free energy decreases monotonously with the increase
of separation (the increase of ${\cal F}/E_C$ is Fig.~\ref{fg5}
is explained by the fact that $|E_C|$ decreases with
separation faster than $|{\cal F}|$). In Fig.~\ref{fg5}(b)
similar results are shown at $T=300\,$K. Here, to avoid an
overlap of the lines, we consider a more narrow separation
region from 50\,nm to $1\,\mu$m. The bottom and top lines
correspond to $\Delta=0.1$ and $\lesssim 0.01\,$eV,
respectively. As can be seen from Fig.~\ref{fg5}(b),
at $T=300\,$K the dependence of the Casimir free energy
on the mass gap parameter of the Dirac model is very weak.
Moreover, at separations above 50\,nm the magnitude of
the free energy ${\cal F}$ decreases with separation as
$a^{-2}$, so that the quantity ${\cal F}/E_C$ is nearly
linear function of separation.

\begin{figure*}[h]
\vspace*{-1.0cm}
\centerline{\hspace*{2.0cm}
\includegraphics{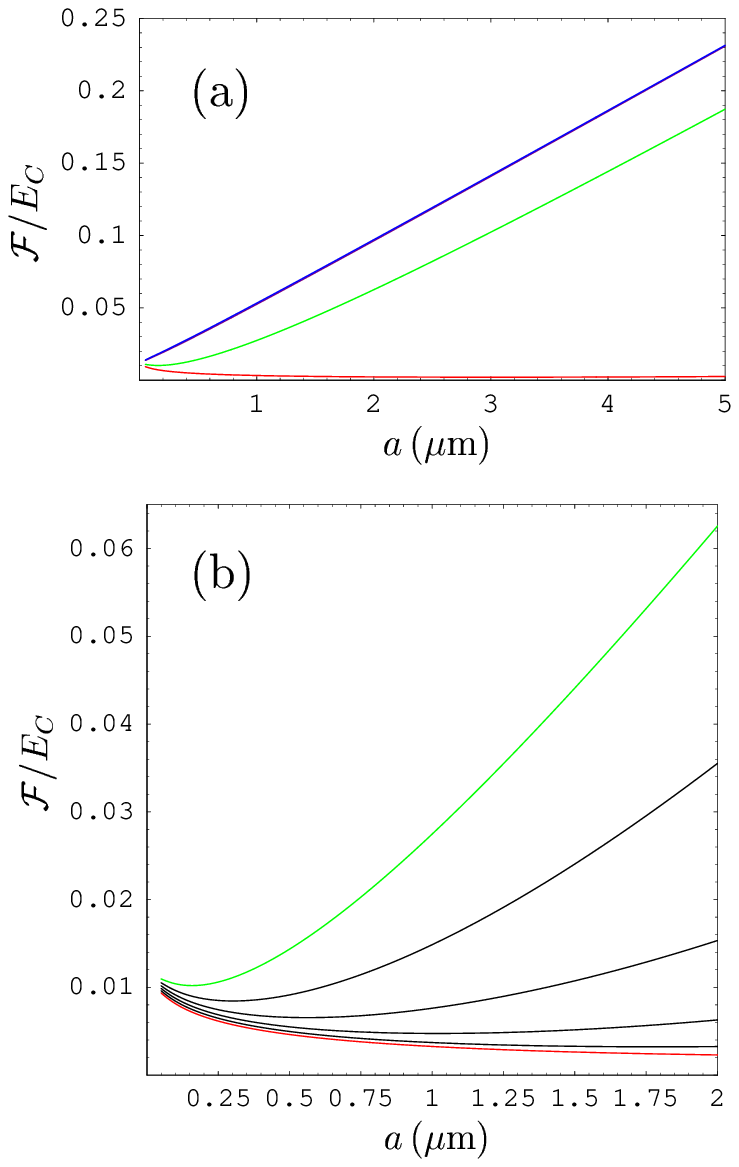}
}
\vspace*{-14.cm}
\caption{\label{fg6}(Color online)
The Casimir free energy as a function of separation
for the interaction of
graphene described by the Dirac model with sapphire plate
at $T=77\,$K
normalized on the Casimir energy between two parallel ideal
metal planes.
The lines from bottom to top correspond to the mass gap
parameter (a) $\Delta=0.1$, 0.05,  $\lesssim 0.01\,$eV
and (b) $\Delta=0.1$, 0.09, 0.08, 0.07, 0.06, $0.05\,$eV,
respectively.
}
\end{figure*}

Similar computations were performed for a graphene
interacting with a sapphire plate. In Fig.~\ref{fg6}(a)
the computational results for ${\cal F}/E_C$
as a function of separation at $T=77\,$K
are presented, where the lines from bottom to top correspond
to  $\Delta= 0.1$, 0.05, and
$\lesssim 0.01\,$eV, respectively.
These results are similar to those presented in Fig.~\ref{fg5}(a)
for a silicon plate. Again, at $T=77\,$K the Casimir free energy
is strongly affected by the value of $\Delta$. To illustrate this
in more detail, in Fig.~\ref{fg6}(b) we plot ${\cal F}/E_C$
over the separation region from 50\,nm to $2\,\mu$m where the
lines from bottom to top correspond to $\Delta=0.1$, 0.09, 0.08,
0.07, 0.06, and 0.05\,eV. At $T=300\,$K, similar to
Fig.~\ref{fg5}(b), the dependence of the free energy on $\Delta$
becomes very weak and one obtains ${\cal F}/E_C$ increasing
nearly linear with the increase of $a$.

\subsection{Asymptotic behavior at high temperature}

The asymptotic behavior of the Casimir free energy at high
temperature (or, equivalently, at large separations) can be
obtained by considering the zero-frequency contribution to
the Lifshitz formula (\ref{eq1}). This corresponds to large
values of the dimensionless parameter $\tau$ introduced
after Eq.~(\ref{eq4}). By putting $l=0$, $\zeta_0=0$ in
Eq.~(\ref{eq4}) one arrives at
\begin{eqnarray}
&&
\tilde{\Pi}_{00}(0,y)=8\alpha y^2\int_{0}^{1}dx\frac{x(1-x)}{\theta}
+\frac{8\alpha}{\tilde{v}_F^2}\int_{0}^{1}dx
\label{eq13} \\
&&~\times\left[\frac{\tau}{2\pi}\ln\left(1+2e^{-\frac{2\pi}{\tau}\theta}
+e^{-\frac{4\pi}{\tau}\theta}\right)+\frac{\tilde{\Delta}^2}{\theta}
\frac{1+e^{-\frac{2\pi}{\tau}\theta}}{1+\cosh\frac{2\pi}{\tau}\theta}
\right],
\nonumber
\end{eqnarray}
\noindent
where
\begin{equation}
\theta\equiv\theta(x,y)=\left[\tilde{\Delta}^2+x(1-x)\tilde{v}_F^2y^2
\right]^{1/2}.
\label{eq14}
\end{equation}
\noindent
Equation (\ref{eq13}) can be identically rearranged to the form
\begin{eqnarray}
&&
\tilde{\Pi}_{00}(0,y)=\frac{8\alpha}{\tilde{v}_F^2}
\left[\frac{\tau}{\pi}\int_{0}^{1}dx\ln\left(2
\cosh\frac{\pi\theta}{\tau}\right)\right.
\nonumber \\
&&~~~~~~~~~~\left.
-\tilde{\Delta}^2
\int_{0}^{1}\frac{dx}{\theta}\tanh\frac{\pi\theta}{\tau}
\right].
\label{eq15}
\end{eqnarray}

In the limit of high temperature we assume that
$\pi\tilde{f}_F/\tau\ll 1$. In this case
\begin{equation}
\frac{\pi\theta}{\tau}\approx\frac{\pi\tilde{\Delta}}{\tau}=
\frac{\Delta}{2k_BT}
\label{eq16}
\end{equation}
\noindent
and Eq.~(\ref{eq15}) is reduced to
\begin{equation}
\tilde{\Pi}_{00}(0,y)\approx\frac{8\alpha}{\tilde{v}_F^2}
\,\frac{\tau}{\pi}\,\ln\left(2
\cosh\frac{\pi\tilde{\Delta}}{\tau}\right)
\equiv\tilde{\Pi}_{00}(0).
\label{eq17}
\end{equation}
\noindent
{}From Eq.~(\ref{eq2}) the TM reflection coefficient on graphene
at zero Matsubara frequency is given by
\begin{eqnarray}
r_{\rm TM}^{(g)}(0,y)&=&
\frac{\tilde{\Pi}_{00}(0,y)}{\tilde{\Pi}_{00}(0,y)+2y}=
1-\frac{2y}{\tilde{\Pi}_{00}(0,y)+2y}
\nonumber \\
&\approx&
1-\frac{2y}{\tilde{\Pi}_{00}(0)}.
\label{eq18}
\end{eqnarray}
\noindent
Taking into account that in accordance with Eq.~(\ref{eq9}) for
dielectric materials $r_{\rm TE}^{(p)}(0,y)=0$, we obtain from
Eqs.~(\ref{eq1}) and (\ref{eq18}) the following asymptotic
expression for the Casimir free energy at large $\tau$:
\begin{equation}
{\cal F}(a,T)\approx\frac{k_BT}{16\pi a^2}\int_{0}^{\infty}
\!\!\!\!\!ydy\ln\left\{1-r_0\left[1-\frac{2y}{\tilde{\Pi}_{00}(0)}
\right]e^{-y}\right\},
\label{eq19}
\end{equation}
\noindent
where the TM reflection coefficient of the dielectric plate at
zero Matsubara frequency
\begin{equation}
r_{\rm TE}^{(p)}(0,y)\equiv r_0=
\frac{\varepsilon(0)-1}{\varepsilon(0)+1}.
\label{eq20}
\end{equation}
\noindent
Equation (\ref{eq20}) can be rearranged to the form
\begin{eqnarray}
&&
{\cal F}(a,T)\approx\frac{k_BT}{16\pi a^2}\int_{0}^{\infty}
\!\!\!ydy\ln\left\{
\vphantom{\left[\frac{2}{\tilde{\Pi}_{00}}\frac{r_0e^{-y}}{r_0e^{-y}}\right]}
(1-r_0e^{-y})\right.
\nonumber \\
&&~~~\times\left.
\left[1+\frac{2y}{\tilde{\Pi}_{00}(0)}
\frac{r_0e^{-y}}{1-r_0e^{-y}}\right]\right\}.
\label{eq21}
\end{eqnarray}
\noindent
In view of the fact that $\tau/(\pi\tilde{v}_F)\gg 1$, one obtains
\begin{equation}
{\cal F}(a,T)\approx\frac{k_BT}{16\pi a^2}\int_{0}^{\infty}
\!\!\!\!\!ydy\left[\ln(1-r_0e^{-y})+\frac{2y}{\tilde{\Pi}_{00}(0)}
\frac{r_0e^{-y}}{1-r_0e^{-y}}\right].
\label{eq22}
\end{equation}
\noindent
Performing the integration in Eq.~(\ref{eq22}) we arrive at the
following asymptotic expression for the Casimir free energy:
\begin{eqnarray}
&&
{\cal F}(a,T)=\frac{k_BT}{16\pi a^2}\left[-{\rm Li}_3(r_0)+
\frac{4}{\tilde{\Pi}_{00}(0)}{\rm Li}_3(r_0)\right]
\label{eq23} \\
&&~~~~
=-\frac{k_BT}{16\pi a^2}{\rm Li}_3(r_0)\left[1-
\frac{\pi\tilde{v}_F^2}{2\alpha\tau\ln\left(2\cosh
\frac{\pi\tilde{\Delta}}{\tau}\right)}\right],
\nonumber
\end{eqnarray}
\noindent
where ${\rm Li}_n(z)$ is the polylogarithm function.

The application region of Eq.~(\ref{eq23}) depends on the specific
values of parameters. Thus, for Si at $T=300\,$K Eq.~(\ref{eq23})
leads to less than 1\% errors in the values of the free energy,
as compared with the results of numerical computations, at $a\geq
500\,$nm for graphene with $\Delta=0.1\,$eV and at $a\geq 1\,\mu$m
for graphene with $\Delta=0.01\,$eV. For Si at $T=77\,$K and
graphene with $\Delta=0.1\,$eV Eq.~(\ref{eq23}) is not yet applicable
at $a=5\,\mu$m and for graphene with $\Delta=0.01\,$eV works
well for $a\geq 4\,\mu$m.

It is interesting to compare the asymptotic expression
(\ref{eq23})
with other results obtained in the literature.
Thus, using the nonlocal dielectric function in the random phase
approximation, the free energy of graphene interacting with a
dielectric substrate (SiO${}_2$) at large separations was
found\cite{21b} to decrease as $a^{-3}$. This is not in
accordance with the main term of
our result (\ref{eq23}) which demonstrates the
classical limit, as is expected at large $a$.
Note that another work\cite{34a} models the dielectric properties
of graphene by the Drude-type function and arrives at the
$a^{-2}$ scaling for graphene-graphene interaction which
satisfies the classical limit.

\section{Thermal interaction of graphene described by the Dirac
model with metallic plate}

The case of graphene interacting with metallic plate is of
special interest. As was mentioned in Sec.~I, there are two
different theoretical approaches to the description of Casimir
effect between real-metal plates. The Drude model approach takes
into account the relaxation properties of conduction electrons.
In the framework of this approach, the imaginary part of the
dielectric permittivity of the Drude model is used to extrapolate
${\rm Im}\,\varepsilon^{\rm opt}(\omega)$, obtained from the
measured optical data, to zero frequency. By contrast, the plasma
model approach disregards relaxation processes and extrapolates
$\varepsilon^{\rm opt}(\omega)$ to zero frequency by means of the
simple plasma model. Although the Drude model approach may seem
preferable, as it takes into account some really existing property
of metals, the experimental situation more likely favors the
plasma model approach. In a series of precise independent
measurements performed by the two experimental
groups\cite{34,35,36,37,38} the Drude model approach was
excluded at a high confidence level (several
experiments\cite{39,40,41,42,43,44} also excluded the influence
 of free charge carriers, that are present
in dielectric materials at room temperature, on the Casimir force).
The two experiments that support the Drude model
approach\cite{45,46} are not independent measurements of the
Casimir force; they are based on fitting procedures between
measured data for the total force and theoretical predictions
using hypothetical models for the electric contribution to it.
Here, we show that in the interaction of graphene described
by the Dirac model with metallic plate the results obtained
are not sensitive to the approach used (either Drude or
plasma). This is, however, not the case when graphene is described
by the hydrodynamic model (see Sec.~V).

\subsection{Free energy as a function of temperature}

Numerical computations of the Casimir free energy per unit area
between graphene described by the Dirac model and Au plate were
performed by using Eqs.~(\ref{eq1}), (\ref{eq2}),
(\ref{eq4})--(\ref{eq6}) and (\ref{eq9}) with $\mu_l=1$.
The dielectric permittivity of Au along the imaginary frequency
axis was described either by the generalized Drude-like
model with temperature-dependent relaxation
parameter\cite{47,48} or by the generalized plasma-like
model.\cite{23,27,37} These models use the six-oscillator
approximation for the optical data extrapolated to zero
frequency by means of simple Drude and plasma models,
respectively, with the plasma frequency $\omega_p=9.0\,$eV
and the relaxation parameter at room temperature
$\gamma=0.035\,$eV. At lower $T$ the lower values of $\gamma$
according to the standard theory of electron-phonon
interaction have been used.\cite{49}

\begin{figure*}[h]
\vspace*{-1.0cm}
\centerline{\hspace*{-1cm}
\includegraphics{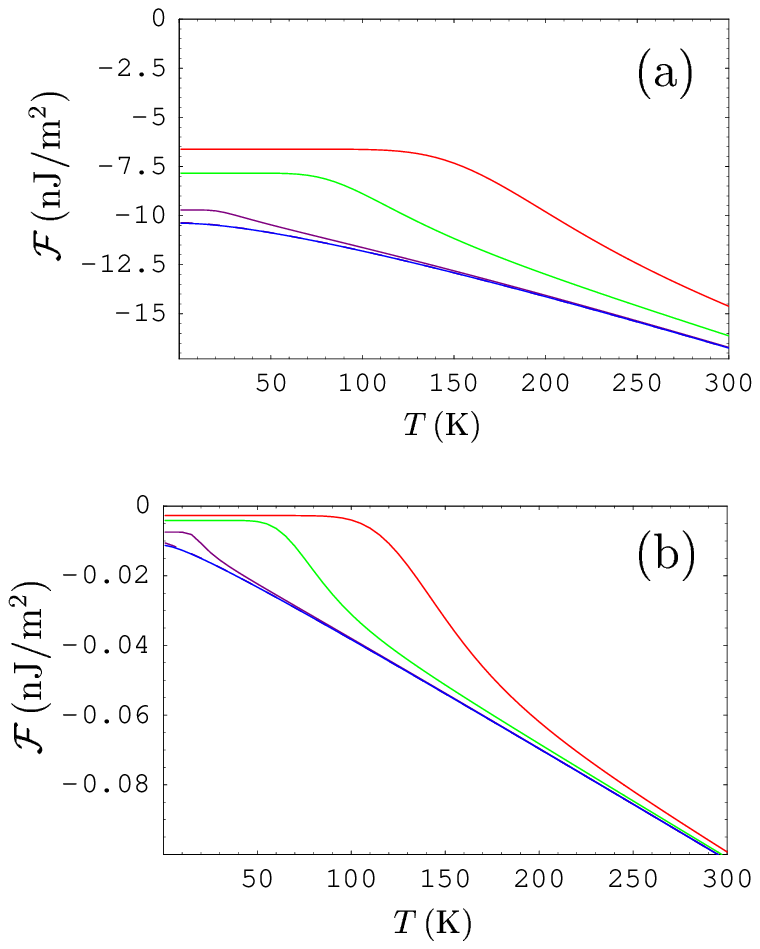}
}
\vspace*{-16.cm}
\caption{\label{fg7}(Color online)
The Casimir free energy per unit area as a function of temperature
for the interaction of
graphene described by the Dirac model with Au plate
(a) at $a=100\,$nm and (b) at $a=1\,\mu$m.
The lines from top to bottom correspond to the mass gap
parameter $\Delta=0.1$, 0.05, 0.01, and $\lesssim 0.001\,$eV,
respectively.
}
\end{figure*}

\begin{figure*}[h]
\vspace*{-1.0cm}
\centerline{\hspace*{2.0cm}
\includegraphics{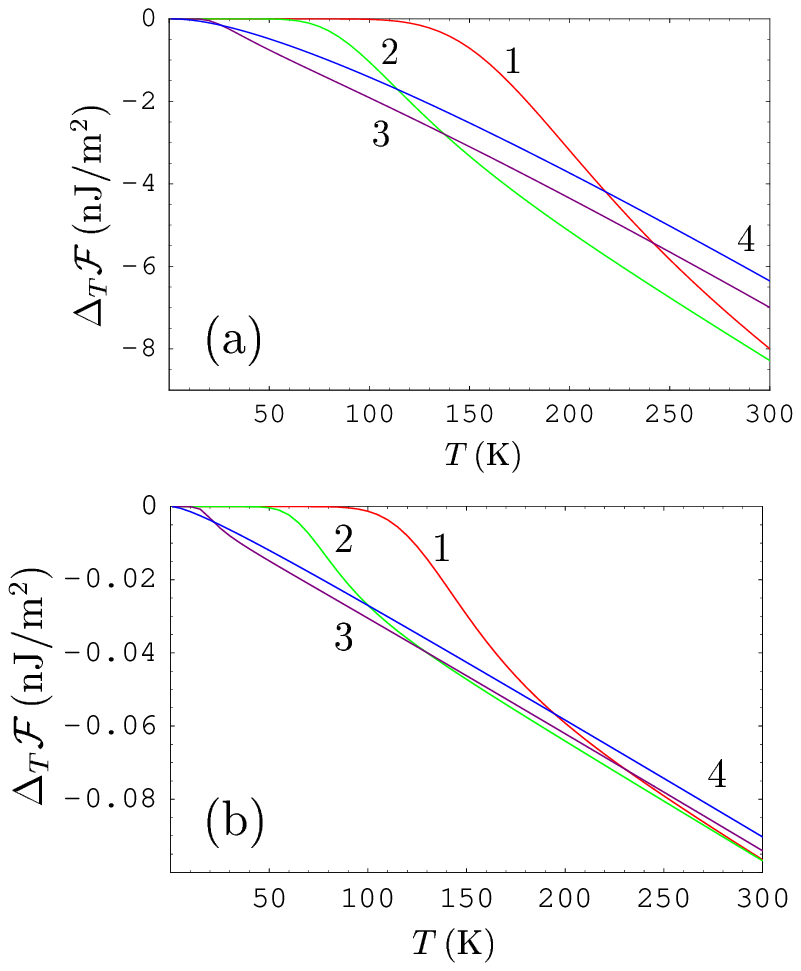}
}
\vspace*{-16.cm}
\caption{\label{fg8}(Color online)
The thermal correction to the Casimir energy per unit area
at zero temperature as a function of temperature
for the interaction of
graphene described by the Dirac model with Au plate
(a) at $a=100\,$nm and (b) at $a=1\,\mu$m.
The lines labeled 1, 2, 3, and 4 correspond to the mass gap
parameter $\Delta=0.1$, 0.05, 0.01, and $\lesssim 0.001\,$eV,
respectively.
}
\end{figure*}

The computational results using the Drude- and plasma-model
approaches
are found to be indistinguishable. Thus, at $T=77\,$K
the relative difference between the Casimir free energies computed
using both approaches achieves the maximum value of 0.02\%
at $a=50\,$nm, does not depend on $\Delta$ in the limits
of our computational accuracy, and decreases with the
increase of separation. At $T=300\,$K this difference
achieves the maximum values of 0.06\% at $\Delta=0$ and
0.07\% at $\Delta=0.1\,$eV, and again decreases with
increasing $a$. Figure \ref{fg7} presents the Casimir free energy
per unit area as a function of temperature (a) for
$a=100\,$nm and (b) for $a=1\,\mu$m.
It can be seen that Fig.~\ref{fg7} demonstrates the same
characteristic features, as Figs.~\ref{fg1} and \ref{fg4}
plotted for dielectric plates, but the magnitudes of the free
energy for the case of metallic plate are larger.
The most important novel qualitative effect, which is found for
both dielectrics and metals, is that for each nonzero $\Delta$
the free energy is nearly unchanged with the increase of $T$
within some temperature interval.
In Fig.~\ref{fg8} we present the computational results for the
absolute thermal correction to the Casimir energy at zero
temperature, defined in Eq.~(\ref{eq10}), as a function of
temperature. The lines 1, 2, 3, and 4 correspond to the values
of the mass gap parameter $\Delta=0.1$, 0.05, 0.01, and
$\lesssim 0.001\,$eV, respectively. This figure is analogous
to Fig.~\ref{fg2} plotted for Si. It demonstrates that for a
metallic plate the thermal correction in the graphene-plate
geometry behaves qualitatively in the same way as for a dielectric
plate, but with slightly larger magnitudes of the thermal
correction.

\begin{figure*}[h]
\vspace*{-1.0cm}
\centerline{\hspace*{-1cm}
\includegraphics{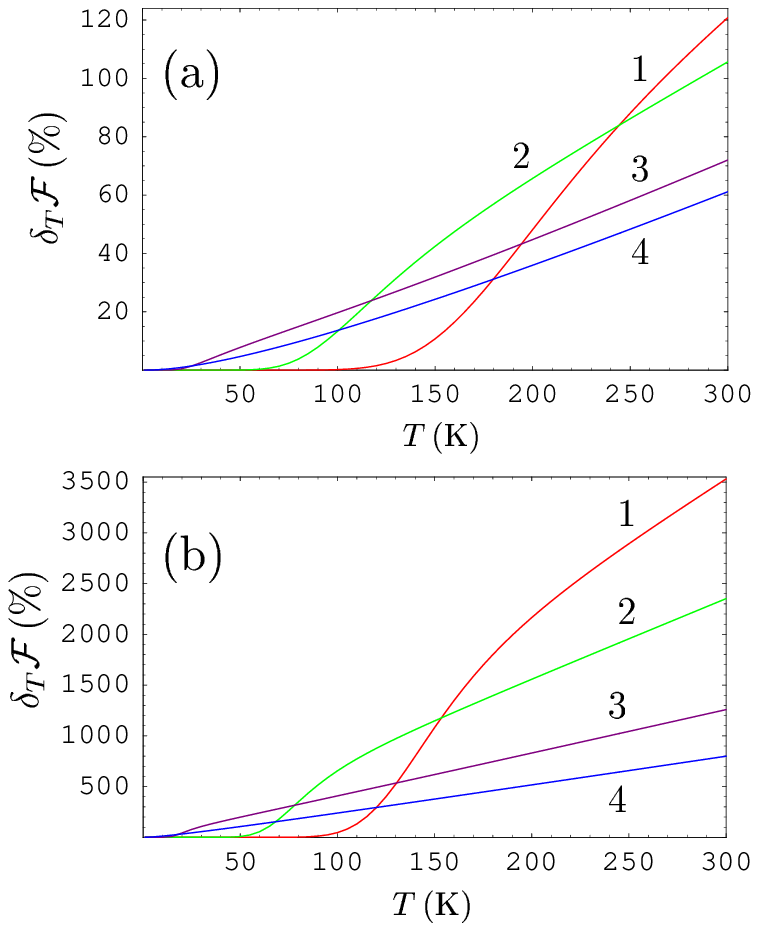}
}
\vspace*{-16.cm}
\caption{\label{fg9}(Color online)
The relative thermal correction to the Casimir energy
at zero temperature as a function of temperature
for the interaction of
graphene described by the Dirac model with Au plate
(a) at $a=100\,$nm and (b) at $a=1\,\mu$m.
The lines labeled 1, 2, 3, and 4 correspond to the mass gap
parameter $\Delta=0.1$, 0.05, 0.01, and $\lesssim 0.001\,$eV,
respectively.
}
\end{figure*}

The computational results for the relative thermal correction,
defined in Eq.~(\ref{eq11}) are presented in Fig.~\ref{fg9}
as a function of temperature. The lines 1, 2, 3, and 4 again
correspond to the same respective $\Delta$, as in Fig.~\ref{fg8}.
Figure \ref{fg9} is analogous to Fig.~\ref{fg3} plotted for a
dielectric plate (silicon). For a metallic plate at $a=100\,$nm
the relative thermal correction at room temperature appears only
slightly larger than for a dielectric plate. At $a=1\,\mu$m
at room temperature the relative thermal correction for Au is
smaller than for Si. This is explained by different values
of the Casimir energy at zero temperature.

\subsection{Free energy as a function of separation}

\begin{figure*}[h]
\vspace*{-1.0cm}
\centerline{\hspace*{2.0cm}
\includegraphics{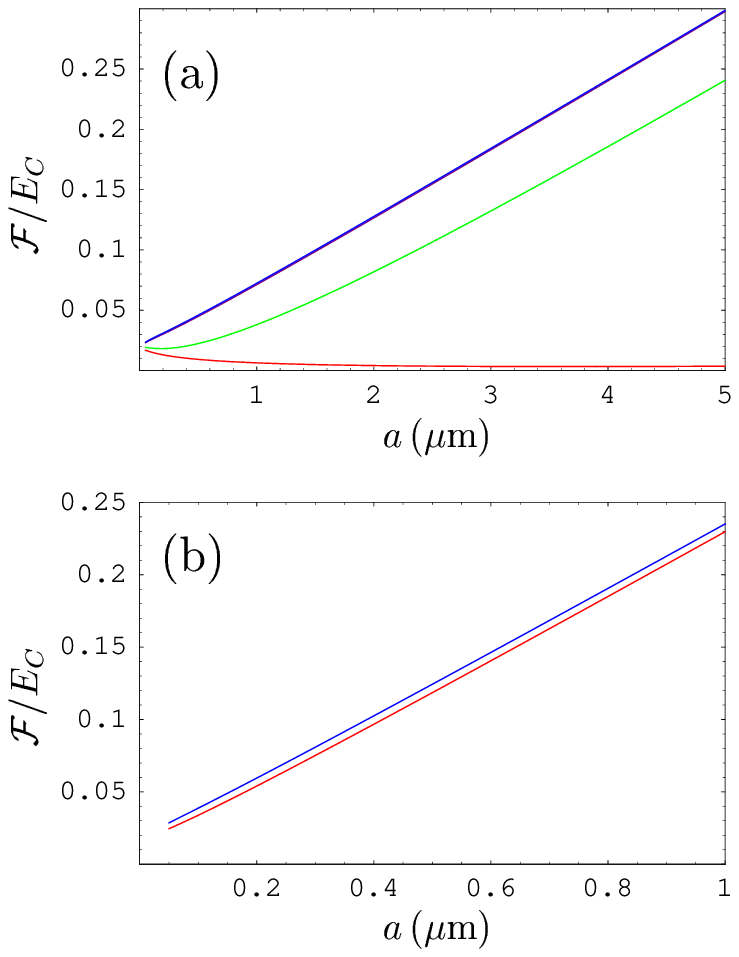}
}
\vspace*{-16.cm}
\caption{\label{fg10}(Color online)
The Casimir free energy as a function of separation
for the interaction of
graphene described by the Dirac model with Au plate
(a) at $T=77\,$K and (b) at $T=300\,$K
normalized on the Casimir energy between two parallel ideal
metal planes.
The lines from bottom to top correspond to the mass gap
parameter (a) $\Delta=0.1$, 0.05,  $\lesssim 0.01\,$eV
and (b) $\Delta=0.1$, $\lesssim 0.01\,$eV,
respectively.
}
\end{figure*}

Keeping in mind that in most experiments on the Casimir force
the temperature is preserved constant and measurements are
performed at different separation distances, here we present
the computational results for a free energy of graphene-metal
interaction as a function of separation. In Fig.~\ref{fg10}(a) the
free energy of graphene interacting with Au plate
normalized on the Casimir energy between ideal metal
planes (\ref{eq12}) is shown. The three lines from bottom to
top correspond to the values of mass gap parameter
$\Delta=0.1$, 0.05, and $\lesssim 0.01\,$eV, respectively.
The obtained values
of the free energy are larger than for graphene interacting with
Si plate [compare with Fig.~\ref{fg5}(a)]. At $T=300\,$K the
dependence of the computational results on the mass gap parameter
becomes not so pronounced as in Fig.~\ref{fg10}(a). This can be
seen in Fig.~\ref{fg10}(b) where the bottom and top lines
correspond to $\Delta=0.1\,$eV and $\lesssim 0.01\,$eV,
respectively.

It is interesting to consider the Casimir interaction of graphene
with a ferromagnetic metal. It was shown\cite{50,51}
that at room temperature the
ferromagnetic properties of real metals may influence the Casimir
force only through the contribution of the zero-frequency term of
the Lifshitz formula. Recently the gradient of the Casimir force
between a nonmagnetic Au sphere and a magnetic metal (Ni) plate
has been measured.\cite{52}
We have computed the Casimir free energy per unit area between a
graphene described by the Dirac model and Ni plate using the same
formalism, as for an Au plate. The dielectric permittivity of Ni
along the imaginary frequency axis was found from the tabulated
optical data\cite{53} extrapolated to zero frequency either by
the Drude or by the plasma model with the plasma frequency
$\omega_p=4.89\,$eV and the relaxation parameter at room
temperature $\gamma=0.0436\,$eV.\cite{53,54}
The value of $\mu(0)=110$ for the static magnetic permeability
of Ni has been used. It was found that relative differences in
the computational results for the free energy of graphene-Ni
interaction, when Ni is described using the Drude- and
plasma-model approaches, are as small as computed above for the
interaction of graphene with an Au plate. The influence of
magnetic properties on the free energy was also shown to be
negligibly small. The relative difference between the free
energies of graphene-Ni and graphene-Au interactions for
graphene with $\Delta=0.1\,$eV computed at $T=300\,$K is
equal to 6\% at $a=100\,$nm and decreases to 1\% at
$a=1\,\mu$m. It is less for smaller values of the mass gap
parameter. Note that even these small differences are not due to
magnetic properties of Ni but due to different plasma frequency
and optical properties of Ni as compared to Au.

\subsection{Asymptotic behavior at high temperature}

Now we derive the analytic expression for the Casimir free energy
of graphene described by the Dirac model interacting with
metallic plate at $\tau\gg 1$. The contribution of the TM
reflection coefficient for graphene interacting with dielectric
plate was obtained in Eq.~(\ref{eq22}). Taking into account that
for metallic materials $r_0$ defined in Eq.~(\ref{eq20}) is equal
to unity, one obtains from Eq.~(\ref{eq22})
\begin{equation}
{\cal F}_{\rm TM}(a,T)\approx\frac{k_BT}{16\pi a^2}\int_{0}^{\infty}
\!\!\!\!ydy\left[\ln(1-e^{-y})+\frac{2y}{\tilde{\Pi}_{00}(0)}
\frac{e^{-y}}{1-e^{-y}}\right].
\label{eq24}
\end{equation}
\noindent
Calculating the integral with respect to $y$ and using Eq.~(\ref{eq17}),
 we arrive at
\begin{eqnarray}
&&
{\cal F}_{\rm TM}(a,T)=-\frac{k_BT}{16\pi a^2}\zeta(3)\left[1-
\frac{4}{\tilde{\Pi}_{00}(0)}\right]
\label{eq25} \\
&&~~
\approx
-\frac{k_BT}{16\pi a^2}\zeta(3)\left[1-
\frac{\pi\tilde{v}_F^2}{2\alpha\tau\ln\left(2\cosh
\frac{\pi\tilde{\Delta}}{\tau}\right)}\right],
\nonumber
\end{eqnarray}
\noindent
where $\zeta(z)$ is the Riemann zeta function.
This result in the special case $\tilde{\Delta}=0$ was obtained in
Ref.~\cite{55}. Note that for a dielectric plate considered in
Sec.~IIIC the contribution of the TM mode was in fact equal to the
total free energy because $r_{\rm TE}^{(p)}(0,y)=0$ for
dielectrics.
For metals this is in general not so (see below).

The TE reflection coefficient for graphene at zero Matsubara
frequency is obtained from Eq.~(\ref{eq2})
\begin{equation}
r_{\rm TE}^{(g)}(0,y)=
-\frac{\tilde{\Pi}_{tr}(0,y)-
\tilde{\Pi}_{00}(0,y)}{\tilde{\Pi}_{tr}(0,y)-\tilde{\Pi}_{00}(0,y)+2y}.
\label{eq26}
\end{equation}
\noindent
{}From Eq.~(\ref{eq6}) taken at $l=0$, $\zeta_0=0$ and Eq.~(\ref{eq13})
it is easily seen that
\begin{eqnarray}
&&
\tilde{\Pi}_{tr}(0,y)-\tilde{\Pi}_{00}(0,y)=8\alpha\tilde{v}_F^2y^2
\int_{0}^{1}dx\frac{x(1-x)}{\theta}
\nonumber \\
&&~~~~~~~\times
\left(1-\frac{1+e^{-\frac{2\pi}{\tau}\theta}}{1+\cosh\frac{2\pi}{\tau}\theta}
\right),
\label{eq27}
\end{eqnarray}
\noindent
where the quantity $\theta$ is defined in Eq.~(\ref{eq14}). After
identical transformations with account of Eq.~(\ref{eq16}) the
result is
\begin{eqnarray}
&&
\tilde{\Pi}_{tr}(0,y)-\tilde{\Pi}_{00}(0,y)=8\alpha\tilde{v}_F^2y^2
\int_{0}^{1}dx\frac{x(1-x)}{\theta}
\tanh\frac{\pi\theta}{\tau}
\nonumber \\
&&~~~~~
\approx\frac{8\alpha\tilde{v}_F^2y^2}{\tilde{\Delta}}
\tanh\frac{\pi\tilde{\Delta}}{\tau}
\int_{0}^{1}x(1-x)dx
\nonumber\\
&&~~~~~=
\frac{4\alpha\tilde{v}_F^2y^2}{3\tilde{\Delta}}
\tanh\frac{\pi\tilde{\Delta}}{\tau}.
\label{eq28}
\end{eqnarray}
\noindent
This quantity is negligibly small as compared to unity because the
main contribution to the Lifshitz formula (\ref{eq1}) is given by
$y\sim 1$ and for $\tilde{\Delta}\to 0$ one has
\begin{equation}
\tilde{\Pi}_{tr}(0,y)-\tilde{\Pi}_{00}(0,y)\to
\frac{4\pi\alpha\tilde{v}_F^2y^2}{3\tau}.
\label{eq29}
\end{equation}
\noindent
Thus, we can neglect by the difference of polarization operators
in the denominator of Eq.~(\ref{eq26}) and get
\begin{equation}
r_{\rm TE}^{(g)}(0,y)\approx -
\frac{2\alpha\tilde{v}_F^2y}{3\tilde{\Delta}}
\tanh\frac{\pi\tilde{\Delta}}{\tau}.
\label{eq30}
\end{equation}
\noindent
Using the Lifshitz formula (\ref{eq1}) and Eq.~(\ref{eq30}) for
the contribution of the TE reflection coefficient to the Casimir
free energy of graphene-metal interaction at high temperature, we
arrive at
\begin{eqnarray}
&&
{\cal F}_{\rm TE}(a,T)\approx\frac{k_BT}{16\pi a^2}\int_{0}^{\infty}
\!\!\!\!\!ydy\ln\left[1-r_{\rm TE}^{(g)}(0,y)r_{\rm TE}^{(p)}(0,y)e^{-y}\right]
\nonumber \\
&&~~~\approx
\frac{k_BT}{24\pi a^2}\,\frac{\alpha\tilde{v}_F^2}{\tilde{\Delta}}
\tanh\frac{\pi\tilde{\Delta}}{\tau}
\int_{0}^{\infty}\!\!\!y^2dy\,r_{\rm TE}^{(p)}(0,y)e^{-y}.
\label{eq31}
\end{eqnarray}
\noindent
Here we have used that $|r_{\rm TE}^{(g)}(0,y)|\ll 1$ at
$y\sim 1$.

Now we are in a position to consider metallic plates made of
nonmagnetic and magnetic metals described within both the Drude
and the plasma model approaches and in all cases find the
high-temperature behavior of the total Casimir free energy.
We begin with a nonmagnetic metal described by the Drude model
approach. In this case from Eq.~(\ref{eq9}) one obtains that
$r_{\rm TE}^{(p)}(0,y)=0$ and in accordance with Eq.~(\ref{eq31})
the TE contribution to the free energy ${\cal F}_{\rm TE}(a,T)$
vanishes. Thus, for the plate made of a nonmagnetic Drude metal the
total free energy of graphene-metal interaction at high
temperature is given by Eq.~(\ref{eq25}).

Next we consider a nonmagnetic metal described by the
plasma-model approach. In this case from Eq.~(\ref{eq9})
we have
\begin{equation}
r_{\rm TE}^{(p)}(0,y)=\frac{\delta y-
\sqrt{1+\delta^2y^2}}{\delta y+
\sqrt{1+\delta^2y^2}}
\approx-(1-2\delta y+2\delta^2y^2),
\label{eq32}
\end{equation}
\noindent
where the parameter $\delta$ is defined as
\begin{equation}
\delta\equiv\frac{\omega_c}{\omega_p}=\frac{c}{2a\omega_p}=
\frac{\delta_p}{2a}\ll 1
\label{eq33}
\end{equation}
\noindent
and $\delta_p\equiv c/\omega_p$ is the effective penetration
depth of electromagnetic oscillations into the metal.
Substituting Eq.~(\ref{eq32}) in Eq.~(\ref{eq31}) and
integrating with respect to $y$, one obtains
\begin{equation}
{\cal F}_{\rm TE}(a,T)\approx -\frac{k_BT}{12\pi a^2}\,
\frac{\alpha\tilde{v}_F^2}{\tilde{\Delta}}
\tanh\frac{\pi\tilde{\Delta}}{\tau}(1-6\delta+24\delta^2).
\label{eq34}
\end{equation}
\noindent
The total asymptotic expression for the free energy at high
temperature is given by the sum of (\ref{eq25}) and (\ref{eq34}).
Note that the contribution of the TE mode (\ref{eq34}) is a
negligibly small correction because
$\alpha\tilde{v}_F^2\sim 10^{-7}$.

We are coming now to the consideration of magnetic metals
described by the Drude model. {}From Eq.~(\ref{eq9}) it
follows
\begin{equation}
r_{\rm TE}^{(p)}(0,y)=\frac{\mu(0)-1}{\mu(0)+1}.
\label{eq35}
\end{equation}
\noindent
The substitution of this reflection coefficient in
Eq.~(\ref{eq31}) results in
\begin{eqnarray}
&&
{\cal F}_{\rm TE}(a,T)\approx
\frac{k_BT}{24\pi
a^2}\,\frac{\alpha\tilde{v}_F^2}{\tilde{\Delta}}\,
\tanh\frac{\pi\tilde{\Delta}}{\tau}\,\,
\frac{\mu(0)-1}{\mu(0)+1}
\int_{0}^{\infty}\!\!\!y^2e^{-y}dy
\nonumber \\
&&~~~~~~~~
=\frac{k_BT}{12\pi a^2}\,\frac{\alpha\tilde{v}_F^2}{\tilde{\Delta}}
\tanh\frac{\pi\tilde{\Delta}}{\tau}\,\,
\frac{\mu(0)-1}{\mu(0)+1}.
\label{eq36}
\end{eqnarray}
\noindent
This term is again negligibly small, as compared to
${\cal F}_{\rm TM}(a,T)$, so that the total Casimir free energy at
high temperature is well described by Eq.~(\ref{eq25}).

Finally we consider the asymptotic expression for the free energy
of graphene interacting with a magnetic metal described by the
plasma model. In this case from Eq.~(\ref{eq9}) we get
\begin{eqnarray}
&&
r_{\rm TE}^{(p)}(0,y)=\frac{\delta\sqrt{\mu(0)} y-
\sqrt{1+\frac{\delta^2y^2}{\mu(0)}}}{\delta\sqrt{\mu(0)} y+
\sqrt{1+\frac{\delta^2y^2}{\mu(0)}}}
\label{eq37} \\
&&~~~~~~~~~
\approx -\left[1-2\delta\sqrt{\mu(0)}y+2\delta^2\mu(0)y^2
\right].
\nonumber
\end{eqnarray}
\noindent
This is similar to Eq.~(\ref{eq32}) with the replacement
of $\delta$ for $\delta\sqrt{\mu(0)}$. Thus, instead of
Eq.~(\ref{eq34}), one obtains
\begin{eqnarray}
&&
{\cal F}_{\rm TE}(a,T)\approx -\frac{k_BT}{12\pi a^2}\,
\frac{\alpha\tilde{v}_F^2}{\tilde{\Delta}}
\tanh\frac{\pi\tilde{\Delta}}{\tau}
\nonumber \\
&&~~~~~~~~~~\times
\left[1-6\delta\sqrt{\mu(0)}+24\delta^2\mu(0)\right].
\label{eq38}
\end{eqnarray}
\noindent
The total asymptotic expression for the free energy at high
temperature is given by the sum of Eq.~(\ref{eq25}) and
negligibly small addition (\ref{eq38}) depending on the properties
of magnetic metal. The obtained analytic expressions were found in
good agreement with the results of numerical computations within
appropriate temperature (separation) intervals.

\section{Comparison between hydrodynamic and Dirac models of graphene}

On this section we compare the computational results for the free
energy of graphene-plate interaction obtained using two different
models of graphene discussed in Secs.~I and II. We find separation
regions where the predictions of both models are distinct and
similar and compare respective asymptotic expressions for the free
energy at high temperature (large separations).

\subsection{Comparison between computational results for graphene
described by two different models}

\begin{figure*}[h]
\vspace*{-1.0cm}
\centerline{\hspace*{-1cm}
\includegraphics{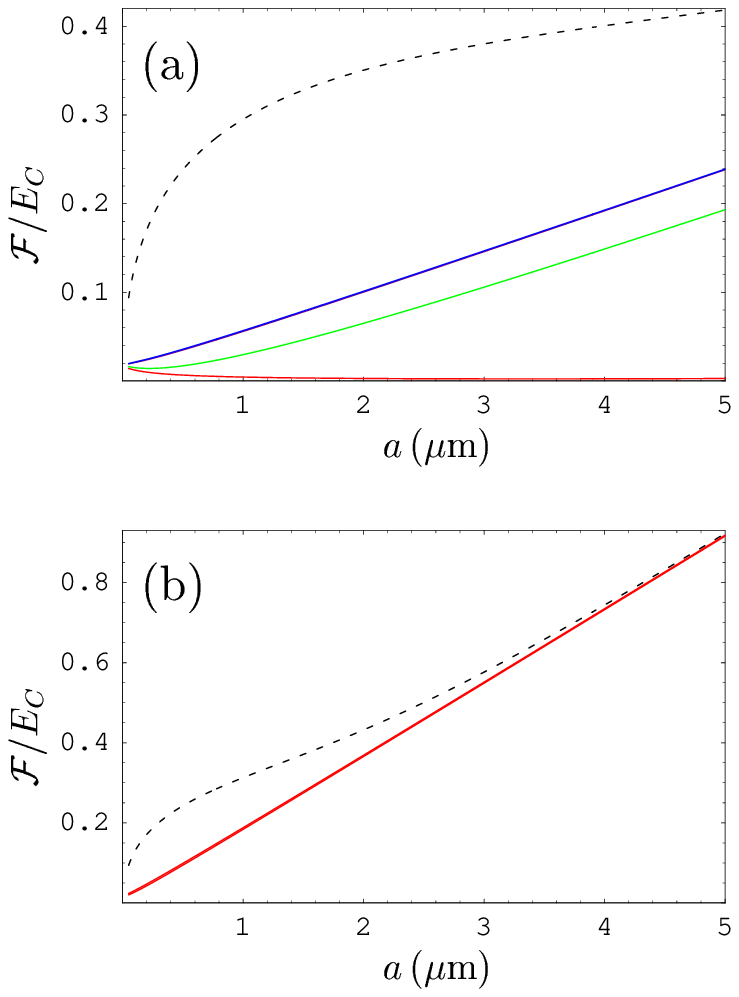}
}
\vspace*{-16.cm}
\caption{\label{fg11}(Color online)
The Casimir free energy as a function of separation
for the interaction of
graphene described by the hydrodynamic model (dashed lines) and
by the Dirac model (solid lines) with Si plate
(a) at $T=77\,$K and (b) at $T=300\,$K
normalized on the Casimir energy between two parallel ideal
metal planes.
The solid lines correspond (a) from bottom to top to the mass gap
parameter $\Delta=0.1$, 0.05,  $\lesssim 0.01\,$eV,
respectively, and (b) to $\Delta=0.1\,$eV.
}
\end{figure*}

Keeping in mind the possibility to compare theoretical predictions
for the Casimir force with the experimental data, we calculate the
free energy of graphene-plate interaction as a function of
separation for both dielectric and metallic plates. Computations
were performed using the Lifshitz formula (\ref{eq1}) where the
reflection coefficients (\ref{eq17}) for graphene in the framework
of the hydrodynamic model were used. The computational results for
${\cal F}/E_C$ as a function of separation are presented in
Fig.~\ref{fg11} by the dashed lines (a) at $T=77\,$K and (b) at
$T=300\,$K. In the same figure the respective results for
${\cal F}/E_C$  computed using the Dirac model of graphene are
reproduced from Fig.~\ref{fg5}(a,b) by the solid lines.
The solid lines in Fig.~\ref{fg11}(a) from bottom to top
correspond to $\Delta=0.1$, 0.05, and $\lesssim 0.01\,$eV,
respectively. Note that in the scale used in Fig.~\ref{fg11}(b)
the two lines of Fig.~\ref{fg5}(b) overlap. They are shown as a
single solid line in Fig.~\ref{fg11}(b).

As is seen in Fig.~\ref{fg11}(a), at $T=77\,$K the hydrodynamic
model of graphene predicts much larger magnitudes of the Casimir
free energy than the Dirac model. Thus, at $T=77\,$K the
predictions of the hydrodynamic model for $|{\cal F}|$ at
$a=0.5\,\mu$m is by factors of 36.0 and 6.7 larger than
predictions of the Dirac model with $\Delta=0.1\,$eV and
$\Delta\lesssim 0.01\,$eV, respectively. At $a=1.5\,\mu$m the
respective factors are 97.9 and 4.2. At $T=300\,$K [see
Fig.~\ref{fg11}(b)] the predictions of the hydrodynamic model
are larger than the predictions of the Dirac model by the
factors of 2.5 and 1.3 at $a=0.5\,\mu$m and $a=1.5\,\mu$m,
respectively. As is seen in Fig.~\ref{fg11}(b), at $T=300\,$K,
$a>4\,\mu$m the asymptotic regime of large $\tau$ is already
achieved and the predictions of the hydrodynamic and Dirac
models almost coincide. At $T=77\,$K [Fig.~\ref{fg11}(a)]
the asymptotic regime of large $\tau$ is achieved at much
larger separations than those shown in the figure.

Now we compare the predictions of the hydrodynamic and Dirac
models of graphene interacting with a metallic plate.
All computations were performed using the same formalism as
above. We considered the plates made of a nonmagnetic metal Au and
a magnetic metal Ni. Each of these metals was described either
using the Drude- or the plasma-model approach.

The computational results for the normalized free energy
${\cal F}/E_C$  are presented in Fig.~\ref{fg12} at $T=300\,$K
as a function of separation. In this figure, the solid line is
reproduced from Fig.~\ref{fg10}(b) [note that in the scale of
Fig.~\ref{fg12} the two solid lines in Fig.~\ref{fg10}(b)
overlap]. This line is obtained using the Dirac model of
graphene interacting with a metallic plate, be it Au or Ni
(see Sec.~IVB). The dashed and dotted lines 1, 2, 3, and 4 are
obtained from computations using the hydrodynamic model of
graphene. The computational data shown by the dashed lines 1
and 3 were found for an Au plate described by the plasma- and
Drude-model approaches, respectively. The data shown by the
dotted lines 2 and 4 were computed for a Ni plate also
described by the respective plasma- and
Drude-model approaches. As can be seen in Fig.~\ref{fg12},
the free energy of graphene-metal interaction with graphene
described by the hydrodynamic model strongly depends on the
metal used (magnetic or nonmagnetic) and on the chosen
approach to the description of dielectric properties.
Only for the nonmagnetic metal (Au) described by the Drude
model (line 3) the behavior of the free energy is qualitatively
similar to the case of dielectric plate and nears the
prediction of the Dirac model at large separations [compare
with Fig.~\ref{fg11}(b)].

As an example, the Casimir free energy of graphene-metal
interaction computed using the hydrodynamic model of graphene at
$a=1\,\mu$m, $T=300\,$K is larger than the same quantity
computed using the Dirac model by factors of 2.37
(for an Au plate described by the plasma model), 2.15
(for a Ni plate described by the plasma model), 1.93
(for an Au plate described by the Drude model), and 1.55
(for a Ni plate described by the Drude model).
This allows comparison between different theoretical
predictions and experimental data. It is interesting that
for lines 1--3 the magnitude of ${\cal F}$ predicted by the
hydrodynamic model is always larger than for the predictions
of the Dirac model (solid line). As to the line 4 (Ni described
by the Drude model), the prediction for $|{\cal F}|$ from the
hydrodynamic model becomes less than from the Dirac model at
$a\approx 1.55\,\mu$m and remains so at larger separations.
Thus, at $a=5\,\mu$m the ratio between the predictions of
hydrodynamic and Dirac models is equal to 0.42.

\subsection{Asymptotic behavior at high temperature}

As was mentioned in Sec.~IIIC, at high temperature (large
separations) the zero-frequency term of the Lifshitz formula
(\ref{eq1}) alone determines the Casimir free energy.
The reflection coefficients on the graphene, described by the
hydrodynamic model, at zero frequency follow from Eq.~(\ref{eq7})
\begin{equation}
r_{\rm TM}^{(g)}(0,y)=1, \qquad
r_{\rm TE}^{(g)}(0,y)=-\frac{\tilde{K}}{\tilde{K}+y}.
\label{eq39}
\end{equation}
\noindent
Taking into account that for a hydrodynamic model at $T=300\,$K
the high-temperature regime starts at $a>5\,\mu$m, we find from
Eq.~(\ref{eq8}) that at these separations $\tilde{K}>6.75$.
Substituting Eq.~(\ref{eq39}) for the reflection coefficient
$r_{\rm TM}^{(g)}$ into the zero-frequency term of the Lifshitz
formula (\ref{eq1}), for the TM contribution to the free energy at
high temperature one obtains
\begin{equation}
{\cal F}_{\rm TM}(a,T)=\frac{k_BT}{16\pi a^2}
\int_{0}^{\infty}\!\!\!\!\!ydy\ln\left[1-r_{\rm TM}^{(p)}(0,y)
e^{-y}\right].
\label{eq40}
\end{equation}
\noindent
For a dielectric plate, using Eq.~(\ref{eq20}) and integrating in
Eq.~(\ref{eq40}), we get
\begin{equation}
{\cal F}_{\rm TM}(a,T)=-\frac{k_BT}{16\pi a^2}{\rm Li}_3(r_0).
\label{eq41}
\end{equation}
\noindent
As noted in Sec.~IIIC, for a dielectric plate the contribution
of the TE mode to the free energy vanishes. Thus, Eq.~(\ref{eq41})
provides the complete expression for the free energy of
graphene-dielectric interaction at high temperature.
Equation (\ref{eq41}) coincides with the first term in Eq.~(\ref{eq23})
obtained for graphene described by the Dirac model [remind that for
the Dirac model of graphene the asymptotic expression (\ref{eq23})
becomes applicable at much smaller separations at the same room
temperature; see Fig.~\ref{fg11}(b)].

\begin{figure*}[h]
\vspace*{-10.0cm}
\centerline{\hspace*{2.0cm}
\includegraphics{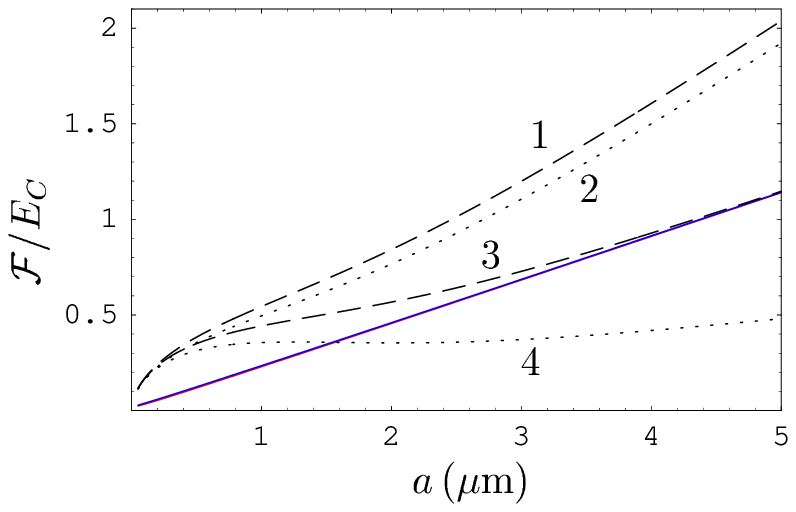}
}
\vspace*{-10.cm}
\caption{\label{fg12}(Color online)
The Casimir free energy as a function of separation
for the interaction of
graphene described by the hydrodynamic model (dashed and dotted lines)
and by the Dirac model (solid line) with Au (dashed lines labeled
1 and 3) and Ni (dotted lines labeled 2 and 4) at $T=300\,$K
normalized on the Casimir energy between two parallel ideal
metal planes.
For lines labeled 1 and 2 metal is described by the plasma-model
approach, and for lines labeled 3 and 4 by the Drude-model
approach.
}
\end{figure*}

For a metallic plate, replacing $r_0$ with unity, we arrive at
\begin{equation}
{\cal F}_{\rm TM}(a,T)=-\frac{k_BT}{16\pi a^2}\zeta(3).
\label{eq42}
\end{equation}
\noindent
This result coincides with the first term of the asymptotic
expression (\ref{eq25}) obtained for graphene described by the
Dirac model. Equation (\ref{eq42}) provides the total asymptotic
expression for the free energy only in the case when a nonmagnetic
metal of the plate is described by the Drude model (see the dashed
line 3 approaching the solid line in Fig.~\ref{fg12} when the
separation distance increases).

Now we consider the contribution of the TE mode to the free energy
of graphene-metal Casimir interaction. {}From Eq.~(\ref{eq39})
we have
\begin{equation}
r_{\rm TE}^{(g)}(0,y)=-\frac{1}{1+\beta y}\approx
-(1-\beta y+\beta^2y^2),
\label{eq43}
\end{equation}
\noindent
where $\beta=1/\tilde{K}$ takes the maximum value
$\beta_{\max}\approx 0.15$ and decreases with further increase
of separation. Then, for the contribution of the TE mode
to the free energy at high temperature one obtains
\begin{eqnarray}
&&
{\cal F}_{\rm TE}(a,T)\approx\frac{k_BT}{16\pi a^2}
\int_{0}^{\infty}\!\!\!\!\!ydy\ln\left[
\vphantom{r_{\rm TE}^{(p)}}
1+(1-\beta y+\beta^2y^2)\right.
\nonumber \\
&&~~~~~~~~~~~~~~~~~~~~\left.
\times r_{\rm TE}^{(p)}(0,y)
e^{-y}\right].
\label{eq44}
\end{eqnarray}

For a nonmagnetic metal described by the plasma model the
reflection coefficient $r_{\rm TE}^{(p)}(0,y)$ is given by
Eq.~(\ref{eq32}). Substituting Eq.~(\ref{eq32}) in
Eq.~(\ref{eq44}), we find
\begin{eqnarray}
&&
{\cal F}_{\rm TE}(a,T)\approx\frac{k_BT}{16\pi a^2}\left[
\int_{0}^{\infty}\!\!\!\!\!ydy\ln(1-e^{-y})+(\beta +2\delta)
\int_{0}^{\infty}\!\frac{y^2dy}{e^y-1}\right.
\nonumber \\
&&~~\left.
-\frac{\beta^2}{2}
\int_{0}^{\infty}\!\!\!y^2dy\frac{2e^y-1}{(e^y-1)^2}-
2\delta(\beta+\delta)
\int_{0}^{\infty}\!\frac{y^2e^ydy}{(e^y-1)^2}\right]
\label{eq45} \\
&&~~=
\frac{k_BT}{16\pi a^2}\left\{-\zeta(3)+2(\beta+2\delta)\zeta(3)-
\frac{\beta^2}{6}[\pi^2+6\zeta(3)]-
\frac{2\pi^2}{3}\delta(\beta+\delta)\right\}.
\nonumber
\end{eqnarray}
\noindent
Taking into account that at $a>5\,\mu$m it holds
$\beta\gg 4\delta$, a more simple expression is also valid
\begin{eqnarray}
&&
{\cal F}_{\rm TE}(a,T)\approx -\frac{k_BT}{16\pi a^2}
\zeta(3)\left[1-2\beta+\beta^2\left(1+\frac{\pi^2}{6\zeta(3)}
\right)\right]
\nonumber \\
&&~~~~~
\approx -\frac{k_BT}{16\pi a^2}
\zeta(3)\left(1-2\beta+2.368\beta^2\right).
\label{eq47}
\end{eqnarray}
\noindent
By combining Eq.~(\ref{eq42}) and Eq.~(\ref{eq47}), the total
free energy for the interaction of graphene with a nonmagnetic
metal described by the plasma model is obtained
\begin{equation}
{\cal F}(a,T)\approx  -\frac{k_BT}{8\pi a^2}
\zeta(3)\left(1-\beta+1.184\beta^2\right).
\label{eq48}
\end{equation}
\noindent
Note that the main contribution to this free energy is twice that
in Eq.~(\ref{eq25}) related to graphene described by the Dirac
model. This explains different behaviors of the dashed line
labeled 1 and the solid line at the largest separations in
Fig.~\ref{fg12}.

We are coming now to a magnetic metal described by the Drude
model. In this case the reflection coefficient of the plate
at zero frequency is given by Eq.~(\ref{eq35}).
The substitution of Eq.~(\ref{eq35}) in Eq.~(\ref{eq44})
leads to
\begin{eqnarray}
&&
{\cal F}_{\rm TE}(a,T)\approx\frac{k_BT}{16\pi a^2}
\int_{0}^{\infty}\!\!\!\!\!ydy\left[\ln(1+r_{\mu}e^{-y})-\beta
\frac{r_{\mu}ye^{-y}}{1+r_{\mu}e^{-y}}\right.
\nonumber \\
&&~~\left.
+\frac{\beta^2}{2}r_{\mu}
\frac{y^2e^{-y}(2+r_{\mu}e^{-y})}{(1+r_{\mu}e^{-y})^2}\right],
\label{eq49}
\end{eqnarray}
\noindent
where
\begin{equation}
r_{\mu}\equiv\frac{\mu(0)-1}{\mu(0)+1}.
\label{49a}
\end{equation}
\noindent
After the integration in Eq.~(\ref{eq49}) the result is
\begin{eqnarray}
&&
{\cal F}_{\rm TE}(a,T)\approx \frac{k_BT}{16\pi a^2}
\left\{-{\rm Li}_3(-r_{\mu})+2\beta {\rm Li}_3(-r_{\mu})\right.
\nonumber \\
&&~~~\left.
-3\beta^2\left[{\rm Li}_3(-r_{\mu})+{\rm Li}_4(-r_{\mu})
\right]\right\}.
\label{eq50}
\end{eqnarray}
\noindent
By combining Eq.~(\ref{eq42}) and Eq.~(\ref{eq50}), we obtain the
following total free energy for graphene interacting with a
magnetic metal described by the Drude model:
\begin{eqnarray}
&&
{\cal F}(a,T)\approx -\frac{k_BT}{16\pi a^2}
\left\{
\vphantom{\beta^2\left[Li_3(-r_{\mu})\right]}
\zeta(3)+{\rm Li}_3(-r_{\mu})-2\beta {\rm Li}_3(-r_{\mu})\right.
\nonumber \\
&&~~~\left.
+3\beta^2\left[{\rm [Li}_3(-r_{\mu})+{\rm Li}_4(-r_{\mu})\right]\right\}
\label{eq51}
\end{eqnarray}
\noindent
As an example, for Ni $r_{\mu}=0.982$ and Eq.~(\ref{eq51})
takes the form
\begin{equation}
{\cal F}(a,T)\approx  -\frac{k_BT}{16\pi a^2}
\zeta(3)\left(0.262+1.47\beta-4.53\beta^2\right).
\label{eq52}
\end{equation}
\noindent
In is seen that the main contribution to this expression differs
from the main term in Eq.~(\ref{eq25}) obtained for the Dirac
model of graphene. This is reflected also in Fig.~\ref{fg12}
(compare the dotted line labeled 4 and the solid line).

For a magnetic metal described by the plasma model we use the
reflection coefficient (\ref{eq37}) and substitute it in
Eq.~(\ref{eq44}). All calculations are similar to the case
of nonmagnetic metal, but the quantity $\delta$ is replaced for
$\delta\sqrt{\mu(0)}$. Thus, instead of Eq.~(\ref{eq45}), we
obtain
\begin{eqnarray}
&&
{\cal F}_{\rm TE}(a,T)\approx
-\frac{k_BT}{16\pi a^2}\zeta(3)\left\{
\vphantom{\left[\frac{\pi^2}{6\zeta(3)}\right]}
1-2\left[\beta+2\delta\sqrt{\mu(0)}\right]\right.
\label{eq53} \\
&&~~\left.
+\beta^2\left[1+\frac{\pi^2}{6\zeta(3)}\right]+
2\delta\sqrt{\mu(0)}\left[\beta+\delta\sqrt{\mu(0)}\right]
\frac{\pi^2}{3\zeta(3)}\right\}.
\nonumber
\end{eqnarray}
\noindent
 In this equation, however, it is impermissible to neglect by the
 quantity $\delta\sqrt{\mu(0)}$ as compared to $\beta$.
Substituting the numerical values of constants to
Eq.~(\ref{eq53}),
we find
\begin{eqnarray}
&&
{\cal F}_{\rm TE}(a,T)\approx
-\frac{k_BT}{16\pi a^2}\zeta(3)\left\{
1-2\left[\beta+2\delta\sqrt{\mu(0)}\right]
\right.
\nonumber \\
&&~~\left.
+2.368\beta^2+
5.474\delta\sqrt{\mu(0)}\left[\beta+\delta\sqrt{\mu(0)}\right]
\right\}.
\label{eq54}
\end{eqnarray}
\noindent
By combining Eq.~(\ref{eq54}) and Eq.~(\ref{eq42}), we arrive at
the total free energy of graphene interacting with a magnetic
metal described by the plasma model
\begin{eqnarray}
&&
{\cal F}(a,T)\approx
-\frac{k_BT}{8\pi a^2}\zeta(3)\left\{
1-\beta-2\delta\sqrt{\mu(0)}
\right.
\nonumber \\
&&~~\left.
+1.184\beta^2+
2.737\delta\sqrt{\mu(0)}\left[\beta+\delta\sqrt{\mu(0)}\right]
\right\}.
\label{eq55}
\end{eqnarray}
\noindent
The main contribution to Eq.~(\ref{eq55}) is by a multiple two
larger than the main contribution to Eq.~(\ref{eq25}) obtained
for the Dirac model of graphene (compare the dotted line labeled
2 and the solid line in Fig.~\ref{fg12}).

The obtained analytic asymptotic expressions for the free energy
of graphene described using the hydrodynamic model in
graphene-metallic plate geometry is in good agreement with the
results of numerical computations. Thus, for an Au plate
described by the Drude- and plasma-model approaches at
$a=5\,\mu$m, $T=300\,$K, the results of analytic and numerical
calculations differ by 0.35\% and 1.4\%, respectively.
For Ni the same relative differences are equal to 9.3\% and
3\%. Note that relatively large deviation obtained for Ni
plate described by the Drude-model approach is explained by
the fact that in this case at $a=5\,\mu$m the linear
asymptotic regime is not yet achieved (see the dotted line
labeled 4 in Fig.~\ref{fg12}).

To conclude, we emphasize that all results for the Casimir
free energy obtained in this and previous sections are simply
convertable to the Casimir force $F(a,T)$ in the experimentally
relevant configuration of a sphere above a plate used in most
of experiments on measuring the Casimir force. This can be
done by means of the PFA which states that
\begin{equation}
F(a,T)=2\pi R{\cal F}(a,T),
\label{eq56}
\end{equation}
\noindent
where $R$ is the radius of the sphere. In our case, where
${\cal F}(a,T)$ is the Casimir free energy between a graphene
sheet and a material plate, the force $F(a,T)$ defined in
Eq.~(\ref{eq56}) can be considered as the Casimir force acting
between a graphene sheet and a material sphere of radius $R$.
As was mentioned in Sec.~I, the error introduced by the use
of the PFA was recently proved\cite{28,28a,29} to be smaller
than $a/R$ (i.e. or order of 0.1\% for the typical values of
parameters). Keeping in mind that many effects considered above
far exceed 100\%, the use of the PFA in the comparison between
experiment and theory is fully justified.

\section{Conclusions and discussion}

In the foregoing, we have investigated the Casimir free energy
and the thermal correction to the Casimir energy at zero
temperature for a suspended graphene sheet interacting with a
material plate, either dielectric or metallic. In so doing
graphene was described by the fully relativistic Dirac model
with temperature-dependent polarization tensor.
The dielectric properties of the plate were described by the
frequency-dependent dielectric permittivity taking into
account the interband transitions of core electrons.
For a metallic plate both the Drude- and plasma-model
approaches suggested in the literature have been used.

The main novel result obtained for both dielectric and
metallic plates is that for graphene with any nonzero
mass gap parameter $\Delta$ there exists temperature
interval where the Casimir free energy remains nearly constant.
This happens under the condition $k_BT\ll\Delta$, which
should be satisfied with a large safety margin.
If this condition is satisfied, the thermal correction to the
Casimir energy at zero temperature remains negligibly small.
We have also demonstrated that under the condition
$\Delta\lesssim k_BT$ the thermal correction becomes
relatively large. This makes possible large thermal
corrections for a graphene sheet interacting with material
plate at rather low temperature (short separations).

With respect to the interaction with a metallic plate,
it was shown that for graphene described by the Dirac model
the computational results for the free energy are nearly
independent on whether the Drude- or plasma-model approach
to the dielectric permittivity of metal is used.
To a large extent the free energy of graphene interacting
with metallic plate is also independent on whether metal is
nonmagnetic or magnetic if graphene is described by the Dirac
model. In all cases considered (dielectric or metallic plate,
nonmagnetic or magnetic, described by the Drude- or
plasma-model approach) the analytic asymptotic expressions for
the Casimir free energy at high temperature (large
separations) have been obtained and compared with the results
of numerical computations.

The Casimir free energies obtained using the Dirac model of
graphene were compared with those calculated using the
hydrodynamic model. It was shown that at moderate
temperatures (separations) the magnitudes of the free energy
computed using the hydrodynamic model of graphene differ
significantly from that computed using the Dirac model.
This can be used for the experimental test of these models.
At large separations (high temeperature) the theoretical
predictions from both models of graphene nearly coincide
for a dielectric plate and for a nonmagnetic metallic plate
described by the Drude model. For a nonmagnetic metallic
plate described by the plasma model and for a magnetic plate
described by any model the hydrodynamic and Dirac
descriptions of graphene lead to quite different results for
the free energy at large separations (high temperature).
This fact can be also used for the experimental test of
different models. In this respect the investigation of the
interaction between graphene and metamaterials\cite{56} is
also of large interest.
We have also found analytic asymptotic expressions for the free
energy at high temperature (large separations) when graphene
described by the hydrodynamic model interacts with a dielectric
plate or with a plate made of a nonmagnetic or magnetic metal.
In the last two cases both the Drude- and plasma-model
approaches have been used for a description of the dielectric
properties of metal. The calculation results obtained from the
asymptotic expressions were found in a very good agreement
with the results of numerical computations.

\section*{Acknowledgments}
This work was supported by the DFG grant BO\ 1112/21--1.
G.L.K.\ and V.M.M.\ are grateful to the Institute for
Theoretical Physics, Leipzig University, where this work
was performed, for kind hospitality.



\end{document}